\documentclass[showpacs,superscriptaddress]{revtex4}
\usepackage[dvips]{graphicx}
\usepackage{amsmath}
\usepackage{bm}
\usepackage{amsfonts}
\usepackage{longtable}
\usepackage{epsfig}
\usepackage{subfigure}
\usepackage{dcolumn}
\usepackage{bm,color}
\usepackage{txfonts}
\usepackage{times}
\newcommand{\ket}[1]{\left| #1 \right\rangle}

\newcommand{\be}{\begin{equation}}
\newcommand{\ee}{\end{equation}}
\newcommand{\bea}{\begin{eqnarray}}
\newcommand{\eea}{\end{eqnarray}}

\begin{document}
\title{Non-Markovianity and  Clauser-Horne-Shimony-Holt (CHSH)-Bell inequality violation  in quantum dissipative systems}
\author{ A. Thilagam}
\email{thilaphys@gmail.com}
\affiliation{Information Science, Engineering and Environment, 
Mawson Institute,
University of South Australia, Australia
 5095.} 
\author{A. R. Usha Devi}
\email{ushathirthahalli@gmail.com}
\affiliation{Department of Physics, Bangalore University, 
Bangalore-560 056, India}
\affiliation{Inspire Institute Inc., Alexandria, Virginia, 22303, USA.}
\date{\today}
\begin{abstract}
We examine the non-Markovian dynamics  in a multipartite system  of two initially  correlated  atomic qubits, 
each located in a single-mode leaky cavity and interacting with its own bosonic reservoir. We show the dominance of
non-Markovian features, as quantified by the difference
in fidelity of the evolved system  with  its density matrix at an earlier time,
 in three specific two-qubit partitions  associated with the   cavity-cavity and
atom-reservoir density matrices within the same subsystem,
and the cavity-reservoir reduced matrix across the two  subsystems.
The  non-Markovianity in the  cavity-cavity subsystem is seen to be optimized in the vicinity of the exceptional point. 
The CHSH-Bell  inequality computed for various two-qubit partitions  show that
high  non-locality present in a specific subsystem  appears in conjunction with
enhanced   non-Markovian dynamics in  adjacent subsystems.
This is in contrast to the matching existence of  non-locality
and quantum correlations in regions spanned by time $t$ and  the cavity
 decay rate, $\lambda_c$ for select partitions.  We discuss the applicability of these
results to photosynthetic systems.
\end{abstract}
\maketitle

\section{Introduction}
One of the most intriguing feature that underpins the many promising applications \cite{Niel}
  of quantum  theory is that  it cannot be reduced to  local realistic models 
in which  the outcomes of local measurements are
determined in advance by hidden variables \cite{bell}.
Non-locality which is linked to the violation of Bell inequality of any form,
is considered to arise from complementarity or the impossibility of 
 simultaneous joint measurements of  observables. 
The Clauser-Horne-Shimony-Holt (CHSH)  inequality \cite{chsh} is the only extremal Bell inequality with two settings and two outcomes per site, and involves the comparison of  predictions of quantum theories
 with  theories  rooted in local realism. 

Entanglement and non-locality are distinct 
entities which are  based on different frameworks.
 While  entanglement is invariably linked  to the mathematical formalism 
of the abstract Hilbert space, non-locality is founded on  
correlations of variables that, while amenable to experimental verification,
has rather debatable origins due to the  paradoxical, non-Boolean  nature of quantum logic,
with striking inconsistencies with the deterministic structures that underpins classical logic.
The non-objectivity-non-locality issue  invariably leads to inherent
difficulties in formulating  a rigorous definition of non-locality.
Accordingly, the violation of any Bell inequality may imply the presence of
non-locality,  however a non violation does not guarantee the local or nonlocal 
nature of the correlations under study. Interestingly, 
the violation of a Bell inequality does not  imply the existence of
non-locality within the system where Bell test was performed,  but it is a requirement \cite{william}
that  non-locality is observed in the vicinity of the quantum system under observation.
While non-locality  is a feature which reflects entanglement, the  converse is not true i.e., 
not all entangled states reveal non-local features.   In other words, non-locality is a sufficient feature but not necessary to reveal entanglement \cite{busce}.

A  closely associated problem that has yet to be examined in depth  is the role of non-divisible
quantum maps, first proposed  by Sudarshan $et$ $al$. \cite{sudar1} and
examined in other related works  \cite{Niel,bre,choi,lind,Kraus,alicki,sudar2}, on the violation of the CHSH inequality and the association between non-locality and non-Markovianity. The challenges of relating non-Markovian 
dynamics and maps that are not completely positive (CP) and which describe the evolution
of open quantum systems is well known \cite{eis}.
The quantum Markov process as opposed to its non-Markovian counterpart is described by a subset of completely positive intermediate time dynamical maps \cite{usud}, is limited by its strict applicability to 
  weak  system-environment interactions. In a dissipative environment,
 as is the case in this study, non-Hermitian
effects are introduced and possibly  a non-Markovian evolution  involving some back flow of 
information  to the system ensues.
It would be interesting to examine  non-Markovian processes of open quantum systems
with non-divisible maps and which incorporate memory feedback mechanisms within the realm of non-locality.
In a recent work \cite{lipra}, the link between non-Markovian revivals and 
 non-locality for a qubit-oscillator system was examined with results
showing a strong correlation between revivals of non-locality and non-Markovian evolution
dynamics of the studied system,  nevertheless these trends were
noted as not being indicative of a non-Markovian
basis  to changes  during violation of the  CHSH inequality.
In an earlier  work \cite{bella1}, the association between  entanglement and non-locality 
was examined  from a dynamical perspective, and closed relations  were obtained for independent environments.

The main interest in non-Markovian processes arise
from the fact that Markovian dynamics is typically only an approximation, which is no longer valid when considering shorter time-scales and/or stronger system-environment couplings as in light-harvesting
systems where non-Markovianity appears  significant under 
physiological conditions \cite{carupra,reben}.
The  non-Markovian measure has several interpretations with its
evaluation  based on a variety of physical measures depending on the evolution
dynamics of the system under study.
 One popular computable measure was proposed by Breuer and coworkers \cite{breu2009},
where Markovianity is  employed as a characteristic of the  dynamical map $\rho(0)\mapsto\rho(t)=\Phi(t,0)\rho(0)$,
and is associated with the decrease in the trace-distance which quantifies 
the  distinguishability,   $D[\rho_1,\rho_2]$ between two system states,
 $\rho_{1},\ \rho_2$ . The distinguishability measure  does not increase under all completely positive, trace preserving maps, hence  $\sigma=dD[\rho_1,\rho_2,t]/dt$  is negative (positive) when information flows from (to) the system to (from) its environment. Consequently, the increase of trace distance during 
any time intervals is taken as a signature  of the emergence of non-Markovianity  over the period of dynamical evolution,
and has been employed to examine dynamics of interactions in   light-harvesting systems \cite{reben}.
The quantitative estimate of non-Markovianity is based on  the cumulant of the positive information flux
involving  a maximization of all  pairs of initial states to determine the largest amount of information 
that  can be recovered from the environment \cite{breu2009}. 
Wolf  $et$ $al$. \cite{wolf}  introduced an alternative measure of non-Markovianity
based on divisibility, and subsequently Rivas $et$ $al$. \cite {Rivas2010}
 also proposed a measure of non-Markovianity based on deviations from  divisibility 
and the characteristics of quantum correlations of the ancilla component of  an entangled system evolving
under  a trace preserving completely positive quantum channel.  
The subject of equivalence of the two 
measures of non-Markovianity for open two-level systems has been topic of much interest in 
 recent works \cite{hao,chru}, with the results indicating a dependence on the generic features
of the quantum system for the two measures to be reconciled. It is to be noted that both 
 measures are based on  deviations from the continuous, memoryless, completely 
positive semi-group feature of Markovian evolution as is the case in 
other measures introduced recently.

Lu $et$ $al$. \cite {lu} defined another measure of non-Markovianity
using the quantum fisher information flow which quantifies the computable phase 
of  the qubit undergoing decoherence due to its
environment. Recently
the  fidelity of teleportation  \cite{thilchem2} was seen 
to exhibit increased oscillation of fidelities in the  non-Markovian regime
of photosynthetic systems. 
Revivals of quantum correlations can 
exhibit revivals even in absence of back-action from the environment to the system,
this   feature  was linked to the non-Markovian character of the map using a suitable quantifier of 
non-Markovianity in Ref. \cite{bella2}.
 Rajagopal and coworkers \cite{raja} have explored  the generality of the Kraus representation 
for both the Markovian and non-Markovian  quantum evolution,  highlighting the role of the
fidelity measure  $F[\rho(t), \rho(t+\tau)]$ \cite{Fidelity} as being able to track salient properties 
of an  evolved density matrix $\rho(t+\tau)$ as compared to a reference density matrix $\rho(t)$. 
This direct approach offers convenience in gauging  non-Markovian signatures
based on the  fidelity difference $F[\rho(t),\rho(t+\tau)]-F[\rho(0),\rho(\tau)]$.

Our aim is twofold: {\it (a)} to examine the general
relationships between non-Markovianity,  non-locality and 
non-classical correlations \cite{zu} in a multipartite 
system consisting of a  qubit interacting  with a structured reservoir in the form of a cavity
in contact with its own reservoir. This multipartite arrangement provides
different channels of decoherence and dissipation: one that occurs between the qubit and cavity and 
the second, between the cavity and reservoir, {\it (b)} to investigate the  characteristic behavior of
the entities in the vicinity of the exceptional points which are topological defects \cite{Heiss},
which occur when  two eigenvalues of an operator coalesce
when selected system parameters are altered. As a result, 
 two mutually orthogonal  states merge into one self-orthogonal state, 
 resulting in a singularity in the spectrum \cite{Heiss}. 
Using the qubit-cavity-reservoir multipartite arrangement, 
we seek to examine the
links between non-Markovianity, violation of the CHSH-Bell inequality
and non-classical correlations in regions governed by the system parameters within
various two-qubit partitions. 

This  paper is organized as follows. In Section \ref{model} we derive analytical expressions
for the atom-cavity-reservoir model with cavity losses based on
 a phenomenological master equation and  the quantum trajectory approach.
In Section \ref{mult}, the entanglement dynamics of a multipartite system  of two noninteracting subsystems, each 
 consisting of the two-level atom coupled to the cavity which interacts with its reservoir source
is examined, with expressions obtained for several bipartite two-qubit density matrices.
In Section \ref{fide2},  analytical expressions of fidelities for several  two-qubit partitions
near the exceptional point are provided.  Analysis of qualitative features of  
non-Markovianity based on the fidelity difference 
measure of specific subsystem are  analysed in Section \ref{fidemar}.
In Subsection \ref{fidemarC}, qualitative features of non-Markovianity of the cavity-cavity subsystem
obtained in Section \ref{fidemar} is compared with those obtained
using the trace-distance difference measure.
In Section \ref{bellC}, we  make  observations based on the
non-Markovian results in Section \ref{fidemar} and results of non-locality computed using the  CHSH-Bell inequality.
In Section \ref{disC}, we  compare results of  Bell non-locality and 
non-classical correlations for  given partitions, and
  discuss  the applicability of the
results obtained in this study to photosynthetic systems in  Section \ref{lhs}.
Finally,  Section \ref{con}  provides our summary and some conclusions.

\section{Atom-cavity-reservoir model with cavity losses}\label{model}

We  consider a multipartite system of Hamiltonian $\hat{H}_T$ which  consists
of   a two-level atom   coupled to a single-mode leaky cavity $s$, which in turn is
  interacting with its own source of bosonic reservoir $r$  ($\hbar$=1)
 \bea
\label{ham}
    \hat{H}_T&=&\hat{H}_s+\hat{H}_r+\hat{H}_I\,  \\
   \hat{H}_s &=&  \omega_0 \; \sigma_+\; \sigma_-+ \omega_c\, a^\dag a + V (\sigma_-\, a^\dag  + \sigma_+\, a)\, 
\label{hamS} \\
    \hat{H_r} &=& \sum_k \omega_k b^\dag_k b_k, \quad 
    \hat{H_I} =\sum_{k=1}^{N}(\varphi_{k}\hat{a}\hat{b}_{k}^{\dag}+ \varphi_{k}^{*}\hat{b}_{k}\hat{a}^{\dag }),
\label{hamR}
\eea
where  $ \omega_0$ is the atomic resonance frequency, 
$\sigma_+ (\sigma_-)$ denotes the raising (lowering) operator of the atom,  
and  $\hat{a}$ \; $(\hat{a}^{\dag })$ annihilates (creates) a photon with
 frequency $\omega_c$ in the cavity mode. The operator $\hat{b}_{k}$ \; $(\hat{b}^{\dag }_{k})$
annihilates (creates) a photon with frequency $\omega_k$ in $k$-th mode  of the reservoir. 
$V$ is the coupling constant between qubit and cavity and  $ \varphi_{k}$ is the linear coupling 
between the cavity and reservoir.

The density operator $\rho$ of the quantum system associated with the total 
 Hamiltonian, $ \hat{H}_T$ (Eq.~(\ref{ham})) is obtained using   the generalized
Liouville-von Neumann equation  $\frac{d \rho}{dt} = -i {\cal L} \rho$, where
${\cal L}$  generates the map from the initial to final density operators via a Liouville
superoperator $\Phi(t,0)$:   $\rho(0)\mapsto\rho(t)=\Phi(t,0)\rho(0)$. 
 For the simple reversible and non-dissipative situation, and one in which
the state commences in an uncorrelated combination of the atom-cavity subsystem
and reservoir, the reduced dynamics of the  atom-cavity subsystem  can be 
obtained in  the  Kraus representation,
by taking the trace over the reservoir degrees of freedom
\be
\label{kraus}
\rho(t)=\sum_i K_i(t)\rho(0) K_i^\dag(t),
\ee
where the operators $K_i$ constitute the Hilbert space of the subsystem, 
with $\sum_i\, K_i^\dag K_i=I$, and  $I$ is the identity matrix. The map
in Eq.~(\ref{kraus})  projects a  density operator to another positive
density operator, and hence is a positive map.  
The separation of  the total   quantum system into a subsystem of immediate interest, 
and the reservoir environment which is generally considered
 to be in thermal equilibrium forms the basis of Eq.~(\ref{kraus}). 
This is not the case in most practical situations when both
the  Liouville operator and  the map $\Phi(t,0)$ (for negative $t$) are not well defined
as the dynamical evolution of the quantum state of the total system (atom, cavity and reservoir
degrees of freedom) becomes  irreversible and dissipative in nature, making it cumbersome
for realistic modelling of even the atom-cavity subsystem.

Using the projection-operator partitioning method proposed by Feshbach \cite{fesh},
the total Hilbert space of  $ \hat{H}_T$ (Eq.~(\ref{ham}))
is  divided into two orthogonal subspaces generated by  the projection operator,
${\cal P}$ and its complementary projection operator ${\cal Q}=1-{\cal P}$.
This allows the convenient study of the subspace of interest (atom-cavity in our case)
within the total quantum system specified in Eq.~(\ref{ham}).
For our system, ${\cal P}= \sigma_+\; \sigma_-  + a^\dag a $ and
${\cal Q}= \sum_k  b^\dag_k b_k$, such that ${\cal P Q}= {\cal Q P}$=0.
This means that the operators of the atom-cavity subsystem and the reservoir subsystem
(which acts as the dissipative source) commute.
The density operator associated with the qubit-cavity system  is  obtained
using
\be
\label{proj}
\rho_s(t)={\cal P} \; \rho {\cal P},
\ee
where $\rho$ on the RHS of Eq.~(\ref{proj}) is the density operator of the total system.
An excitation present in  the two-level atom $a$ (cavity $c$)
is denoted by $\ket{1}_a$  ($\ket{1}_c$), whilst  $\ket{0}_a$  ($\ket{0}_c$)
 correspond to the ground state atom (vacuum state of the cavity). 
The subspace ${\cal P}$ is  spanned by ($\ket{1}_a, \ket{0}_a, \ket{1}_c, \ket{0}_c$)
while ${\cal Q}$ is  spanned by ($\ket{\mathbf{0}}_r, \ket{\mathbf{1}}_r$).
 The reservoir state $\ket{\mathbf{0}}_r$=$\prod_{k}\ket{0}_{k}$
denotes the vacuum of the reservoir, however its ortho-complement  state, $\ket{\mathbf{1}}_r$
needs further explanation. We note that other than evolving under the action
of the Hamiltonian associated with the  atom-cavity subsystem ($\hat{H_s}$, Eq.~(\ref{hamS})),
the reservoir states also evolve under the reservoir Hamiltonian, $\hat{H_r}$ in Eq.~(\ref{hamR}).
Hence we define  the  collective state of the reservoir  sink, $\ket{\mathbf{1}}_{r}$,
 as a series of superposed terms,
consisting of the state with a single excitation,  $\ket{\mathbf{1}}_{r_i}$=
$\frac{1}{R_0} \sum_{k}\;\varphi_{k} |1_{k}\rangle_{r}$ where $R_0 = 
\sqrt{\sum_k |\varphi_{k}|^2}$, and other states 
orthogonal to it, $\ket{\mathbf{1}}_{r_i}$ (i = 2,3...) as follows
\be
\label{reserS}
\ket{\mathbf{1}}_{r}=\frac{1}{\chi_t}  \sum_{i} N_i \;\ket{\mathbf{1}}_{r_i} 
\ee
where $\chi_t$ is the probability amplitude associated with 
the existence of an excitation in the collective reservoir state, with
the atom and cavity, both remaining in the zero state.
$|1_{k}\rangle_{r}$ denotes an excited oscillator  in
the $k$-th reservoir mode, with oscillators  in all other modes
remaining  in the unexcited state.  

Instead of  a   full microscopic derivation of the system dynamics associated with
the Hamiltonian,  $ \hat{H}_T$ (Eq.~(\ref{ham})),
we incorporate cavity
losses by means of a phenomenological master equation
which  provides 
reliable results when the spectrum of the reservoir is approximately flat,
as justified in a recent work \cite{scala} which compared this model
and a  microscopic system-reservoir interaction model.
 The phenomenological master equation   involving the reduced density matrix, $ \rho_s$
appear as 
\be
\label{pme}
	\frac{\mathrm{d}}{\mathrm{d}t} \rho_s = i( \rho_s \;H_s - H_s \; \rho_s) + \frac{\lambda_a}{2} (2 \sigma_- \rho_S \sigma_+ -
 \sigma_+ \sigma_- \rho_s - \rho_s \sigma_+ \sigma_-) + \frac{\lambda_c}{2} (2 a \rho_s a^\dag - a^\dag a \rho_s -
 \rho_s a^\dag a),
\ee
where $\lambda_a$ ($\lambda_c$ )represents the atomic (cavity) decay rate. The  master equation in Eq.~(\ref{pme})
can be further simplified (setting $\lambda_a$=0) as follows
\bea
\label{pme2}
	\frac{\mathrm{d}}{\mathrm{d}t} \rho_s &=& i( \rho_s \; H'_s - H'_s \; \rho_s) +
 \lambda_c a \rho_s a^\dag, \\ \label{pme3}
H'_s &=& H_s \;- \; i \frac{\lambda_c}{2} \;a^\dag a
\eea
where the second term in the  Hamiltonian $H'_s$ 
 constitutes a decay component which contributes to 
the non-Hermitian features in the system. The last term in Eq.~(\ref{pme2})
denotes the action of the jump operator which forms the basis
of the quantum trajectory approach \cite{traj1,traj2,traj3,traj}.
In this approach, incoherent processes associated with non-Hermitian
terms are incorporated as random quantum jumps 
which results in the collapse of the wavefunction.
The net effect is the transfer of states  from
one subspace to the other, ${\cal P}  \rightarrow  {\cal Q}$.
The density operator  is  obtained by taking an ensemble average
of a range of conditioned operators \cite{traj1} at the select time $t$.
It is important to note that while  Eq.~(\ref{pme2})  describing the dynamics of the atom-cavity system
is Markovian within one subspace, non-Markovian dynamics may arise in the interaction
between the atom (cavity) and reservoir in different subspaces.

 In order to obtain  analytical solutions, we
rewrite  $H'_s$ in Eq.~(\ref{pme3}) (c.f. $\hat{H_s}$, Eq.~(\ref{hamS})) as
\bea
\label{pme4}
H'_s &=&
	\omega_0 \; \sigma_+\; \sigma_-+ (\omega_c + \; \delta - \; i \frac{\lambda_c}{2}) a^\dag a 
+ V (\sigma_-\, a^\dag  + \sigma_+\, a)\, , \\ \label{pme5}
& & \delta -  \; i \frac{\lambda_c}{2} = \int \frac{|\varphi_{k}|^2}{\omega-\omega_k}g(\omega_k) d \omega_k
\eea
where $\delta$ arises due to renormalization and energy integration is based on
the assumption that the  reservoir state energies are closely spaced.
We intentionally ignore detailed parameters, such as spectral density and temperature,
 of the reservoir system for simplicity in analysis of the model under study.
These boson bath parameters  invariably determine  the decay rate $\lambda_c$ 
as is obvious in Eq.~(\ref{pme2}). Moreover  information related to
 the distribution of the phonon bath frequency including the cutoff frequency will 
influence the bath memory time, and accordingly the
 decay rate $\lambda_c$ incorporates leakages between subspaces. The latter processes
are expected to result in a complicated mix of the distinct subspaces of ${\cal P}$  and  ${\cal Q}$,
with bearings on  non-trivial  non-Markovian dynamics between two
subsystems present in the Hilbert space of $ \hat{H}_T$ (Eq.~(\ref{ham})),
this will be revealed in forthcoming Sections.

Considering the resonant condition  $ \omega_0 = \omega_c +  \delta$ and 
the presence of a single initial excitation in each subsystem, we write  
the multipartite state  of the atom-cavity-reservoir as
\be
 \ket{\psi_t}= \ket{1}_a\otimes\ket{0}_c\otimes\ket{\mathbf{0}}_r,
\label{inis}
\end{equation}
Following the quantum trajectory dynamics model \cite{traj1}, 
Eq.~(\ref{inis}) evolves
under the action of the Hamiltonian  in Eq.~(\ref{ham}) as follows
\be
\label{evolve}
    \ket{\psi_t}=\xi_t \ket{1}_a\ket{0}_c \ket{\mathbf{0}}_r+
\eta_t \ket{0}_a \ket{1}_c \ket{\mathbf{0}}_r + \chi_t \ket{0}_t \ket{0}_c \ket{\mathbf{1}}_r, 
\ee
where $|\xi_t|^2$ ($|\eta_t|^2$) is the probability that the excitation is present in the
atom (cavity). The probability amplitude, $\chi_t$, introduced earlier in 
Eq.~(\ref{reserS}) is evaluated using
$|\chi_t|^2=1-|\xi_t|^2-|\eta_t|^2$.

For the initial condition, $\xi_0$=1,  $\eta_0$=0,  the analytical expressions for $\xi_t$ and $\eta_t$
can be obtained  using
\bea
\label{soln}
\frac{d \xi_t}{d t} & =  & - i V \eta_t \, \\ \nonumber
\frac{d \eta_t}{d t} & =  & - i V \xi_t -  \frac{\lambda_c}{2}  \eta_t\,
\eea
To simplify further analysis of the problem, we set  $|\xi_t|^2=1-p$, $|\eta_t|^2=q$ and $|\chi_t|^2=p-q$.  
Here  $q(t)$ is the probability of the atomic qubit exchanging  quantum information with the cavity
at  time $t$, and $1-p(t)$ is the probability that the atomic qubit will remain in its excited
state, provided it  was   in the excited state at $t=$0. 
The exchange mechanism  obviously includes  a dissipative measure, 
$\gamma_d$=$p$-$q$  which is  non-zero in the  event that the reservoir states coupled
to the cavity gets excited. Accordingly, there exists a probability $p-q$ that the atomic qubit decays without 
exciting the cavity states, and we note that in the presence of non-Hermitian exchanges $p > q, \gamma_d >$0.
In general $p(t)$ grows with time, and as will be shown 
later, it can incorporate feedback mechanisms associated with the non-Markovian  dynamics
of the multipartite state. 

 Eq.~(\ref{soln}) yields the analytical forms for $p$, $q$  and appear as
\bea
\label{co1}
p&=&1- e^{-\lambda_c t/2}
\left[\cos{\Omega} t +  \frac{\lambda_c}{4 \Omega}\sin{\Omega} t \right]^2 
\\ \label{co2}
q &=&   e^{-\lambda_c t/2} \frac{V^2}{\Omega^2}\sin^2{\Omega} t,
\eea
where the Rabi frequency $2 \Omega= (4 V^2 -(\frac{\lambda_c}{2})^2)^{1/2}$.
Similar expressions as in  Eq.~(\ref{co1}) can be obtained 
via inversion of a Green's function as 
detailed in an earlier work on the dissipative two-level dimer model \cite{thilaJCP}.
We note the existence of  two 
 regimes, depending on the relation between $V$ and $\lambda_c$. The range where 
$V > \frac{\lambda_c}{4}$ ($V < \frac{\lambda_c}{4}$) applies to the coherent (incoherent) tunneling regime, and at  the exceptional point we obtain
\be
\label{exceptt}
 \Omega = 0, \; V = \frac{\lambda_c}{4}
\ee
At this point , both the coherent and incoherent tunneling regime merge and we obtain
\bea
\label{except}
p &=& 1- \left(1+ \frac{\lambda_c t}{4} \right)^2 e^{-\lambda_c t/2} \\
q &=& \left(\frac{\lambda_c t}{4} \right)^2 e^{-\lambda_c t/2}
\eea
The exceptional point is a  topological defect which is present
  in the vicinity
of a level repulsion \cite{Heiss} and unlike degenerate points,
only one eigenfunction exists at the exceptional point due to the merging of
two eigenvalues. 
The critical temperatures at which exceptional points
occurs in photosynthetic systems was identified recently \cite{thilchem2}. 
In the restricted subspace subspaces of ${\cal P}$ occupied by the atom-cavity system, we note that the
exceptional point in  Eq.~(\ref{exceptt}) may take on a range of values as the cavity decay rate,
$\lambda_c$ is linked to  almost continuous range of reservoir attributes.

\subsection{Entanglement dynamics of a multipartite system incorporating  non-Hermitian terms}\label{mult}
Here we extend the  composite system in Eqs. ~(\ref{evolve}) to examine the entanglement
dynamics in a system of two noninteracting subsystems, each 
 consisting of the two-level atom coupled to the cavity which interacts with its reservoir source.
 To simplify the analysis, we consider that the  subsystems have identical environment with respect to their reservoir characteristics and cavity decay attributes, and assume a multipartite system  in the  initial state 
\be
\label{inistate}
    \ket{\Psi_i}= \left(a \ket{1_{a_1}0_{a_2}}+b \ket{0_{a_1}1_{a_2}}\right)\ket{0_{c_1}0_{c_2}}
    \ket{{{\bf 0}}_{r_1}{{\bf 0}}_{r_2}},
\ee
where $a, b$ are complex parameters, and the cavity $c$ and 
reservoir $r$ are present in their vacuum state. The two subsystems are
labeled as $1$ or $2$.

The composite system, $\ket{\Psi_i}$  evolves as
\bea
\label{mut6}
 a \left [ \xi_t \ket{1}_{a_1} \ket{0}_{c_1} \ket{{\bf 0}}_{r_1} +\eta_t \ket{0}_{a_1}\ket{1}_{c_1} \ket{{\bf 0}}_{r_1} +
\chi_t \ket{0}_{a_1} \ket{0}_{c_1} \ket{{\bf 1}}_{r_1} \right ]
\ket{{0}_{a_2}{0}_{c_2}{\bf 0}_{r_2}}
\\ +
b \left [ \xi_t \ket{1}_{a_2} \ket{0}_{c_2} \ket{{\bf 0}}_{r_2} +\eta_t \ket{0}_{a_2}\ket{1}_{c_2} \ket{{\bf 0}}_{r_2} +
\chi_t\ket{0}_{a_2}\ket{0}_{c_2} \ket{{\bf 1}}_{r_2} \right ]
\ket{{0}_{a_1}{0}_{c_1}{{\bf 0}}_{r_1}}
\eea
By tracing out  the degrees of freedom of the qubits associated with the  cavities $c_1,c_2$, and
reservoirs $r_1,r_2$, the reduced density matrix  of the bipartite two-qubit atomic system  at time $t$ (as determined by $p, q$)  is obtained in the basis $\{\ket{00}, \ket{01}, \ket{10},\ket{11}\}$ as
\be
\label{tqubitm}
 {\rho}_{a_1,a_2}(t)=\left(
\begin{array}{cccc}
p & 0 & 0 & 0  \\
0 & |b|^2(1-p)& a^\star b \;(1-p)& 0 \\
0 &b^\star a\;(1-p) &|a|^2(1-p)& 0 \\
0& 0 & 0 & 0 \\
\end{array}
\right)\\
\ee
Likewise, by tracing out  the degrees of freedom of the qubits associated with $a_1,a_2$, and
$r_1,r_2$ ($c_1,c_2$), the reduced density matrix  of the remaining
bipartite-equivalent partitions associated with the  two-cavity  (two-reservoir) system,  
${\rho}_{c_1,c_2}(t)$ (${\rho}_{r_1,r_2}(t)$)
 can be obtained in the basis $\{\ket{00}, \ket{01}, \ket{10},\ket{11}\}$ by the substitution $(1-p)\rightarrow q$ ($(1-p) \rightarrow \gamma_d$).

For the non-equivalent   atom-cavity partition,
we obtain an associated reduced density matrix   (${\rho}_{a_i,c_i}$) (i=1,2) within the same
subsystem as follows
\be
\label{tresm2}
 {\rho}_{a_i,c_i}(t)=\left(
\begin{array}{cccc}
|a|^2 \gamma_d +b^2& 0 & 0 & 0  \\
  0 & |a|^2q&|a|^2\sqrt{1-p}\sqrt{q}& 0 \\
  0 &|a|^2\sqrt{1-p}\sqrt{q}&|a|^2(1-p)& 0 \\
   0& 0 & 0 & 0 \\
\end{array}
\right)\\
\ee
Likewise the reduced density matrix   (${\rho}_{a_i,r_i}$) associated
with the  atom-reservoir partition is obtained as
\be
\label{tresm3}
 {\rho}_{a_i,r_i}(t)=\left(
\begin{array}{cccc}
|a|^2 q +b^2& 0 & 0 & 0  \\
  0 & |a|^2\gamma_d&|a|^2\sqrt{1-p}\sqrt{\gamma_d}& 0 \\
  0 &|a|^2\sqrt{1-p}\sqrt{\gamma_d}&|a|^2(1-p)& 0 \\
   0& 0 & 0 & 0 \\
\end{array}
\right)\\
\ee
while the reduced density matrix   (${\rho}_{c_i,r_i}$) associated
with the  cavity-reservoir  partition is obtained as
\be
\label{tresm4}
 {\rho}_{c_i,r_i}(t)=\left(
\begin{array}{cccc}
|a|^2 (1-p)+b^2& 0 & 0 & 0  \\
  0 & |a|^2\;\gamma_d&|a|^2\sqrt{q}\sqrt{\gamma_d}& 0 \\
  0 &|a|^2\sqrt{q}\sqrt{\gamma_d}&|a|^2 q& 0 \\
   0& 0 & 0 & 0 \\
\end{array}
\right)\\
\ee
The reduced matrices  (${\rho}_{a_1,r_2}$ ($i \neq j$) associated
with the  atom-reservoir partition across different subsystems
is obtained as
\be
\label{tresm5}
 {\rho}_{a_1,r_2}(t)=\left(
\begin{array}{cccc}
|a|^2(q+\gamma_d)+ |b|^2(1-\gamma_d)& 0 & 0 & 0  \\
  0 & |b|^2\;\gamma_d&a^*b\sqrt{\gamma_d (1-p)}& 0 \\
  0 &a b^*\sqrt{\gamma_d (1-p)}&|a|^2 (1-p)& 0 \\
   0& 0 & 0 & 0 \\
\end{array}
\right)\\
\ee
The remaining  inter-system partitions,
${\rho}_{a_i,c_j}$, ${\rho}_{c_i,r_j}$) ($i \neq j$),
 are  expected to display entanglement 
dynamics and non-locality characteristics not distinctly  different
from those considered above, and therefore will be omitted from further consideration.

\section{Fidelity measure for two-qubit partitions}\label{fide2}

The fidelity measure $0\leq F[\rho(t),\rho(t')] \leq 1$   quantifies the distance between 
 the initial state $\rho(t)$ and the evolved state  $\rho(t')$ at a later time $t'$
and is defined as \cite{Fidelity}
\be
\label{fidelity} 
F[\rho(t),\rho(t')]=\left\{{\rm Tr}\left[\sqrt{\sqrt{\rho(t)}\rho(t')\sqrt{\rho(t)}}\right]\right\}^2, 
\ee
and satisfies the symmetry property, $F[\rho(t),\rho(t')]=F[\rho(t'),\rho(t)]$. 
Based on this definition, the fidelity measure 
$\mathit{F}$  between the  initial state at $t$=0 (for which $p,q,\gamma_d$=0)
and which terminate as   the state in Eqs.(\ref{tqubitm}),  (\ref{tresm2}),  (\ref{tresm3}),  (\ref{tresm4}) or
 (\ref{tresm5})   at a later time $t$  is obtained as
\bea
\label{fide0}
\mathit{F}_1\left({\rho}_{a_1,a_2}(0),{\rho}_{a_1,a_2}(t) \right) &=&1-p,
\quad  \left[ e^{-\frac{t \lambda_c }{2}} \left(1+\frac{t \lambda_c }{4}\right)^2 \right]. \\ \nonumber
\mathit{F}_2 \left( {\rho}_{c_1,c_2}(0), {\rho}_{c_1,c_2}(t) \right) &=& 1-q, \quad \left[ 1-\frac{1}{16} e^{-\frac{t \lambda_c }{2}} t^2 \lambda_c ^2 \right]. \\ \nonumber
\mathit{F}_3 \left( {\rho}_{r_1,r_2}(0), {\rho}_{r_1,r_2}(t)\right) &=& 1-\gamma_d, \quad
\left[ \frac{1}{8} e^{-\frac{t \lambda_c }{2}} (8+ (t  \lambda_c)^2 + 4 t \lambda_c) \right].
\\ \nonumber
\mathit{F}_4\left({\rho}_{a_1,c_1}(0),{\rho}_{a_1,c_1}(t) \right) &=&\left(\sqrt{a^4 (1-p)}+\sqrt{b^4+a^2  b^2 \gamma_d }\right)^2,
\quad \left[ \frac{1}{16} \left(\sqrt{a^4 e^{-\frac{t \lambda_c }{2}} (t \lambda_c + 4)^2}+4 \sqrt{b^2 - \frac{1}{8} a^2 b^2 e^{-\frac{t \lambda_c }{2}} (8+ (t  \lambda_c)^2+4 t \lambda_c)}\right)^2\right].
\\ \nonumber
\mathit{F}_5 \left( {\rho}_{a_1,r_1}(0), {\rho}_{a_1,r_1}(t) \right) &=& \left(\sqrt{a^4 (1-p)}+\sqrt{b^4+a^2  b^2 q }\right)^2,
\quad
 \left[ \left(\sqrt{a^4 e^{-\frac{t \lambda_c }{2}} \left(1 + \frac{t \lambda_c }{4}\right)^2}+\sqrt{b^4+\frac{1}{16} a^2 e^{-\frac{t \lambda_c }{2}}
   t^2 \lambda_c ^2 b^2}\right)^2\right]. \\ \nonumber
\mathit{F}_6 \left( {\rho}_{c_1,r_1}(0), {\rho}_{c_1,r_1}(t)\right) &=& b^2+a^2 (1-p), \quad \left[ 
b^2+a^2 e^{-\frac{t \lambda_c }{2}} \left(1 + \frac{ t \lambda_c }{4}\right)^2 \right].
\label{fide1}
\eea
where the corresponding fidelities near the vicinity of the exceptional point are provided within the square brackets.

\begin{figure}[htp]
  \begin{center}
    \subfigure{\label{fig1a}\includegraphics[width=4.5cm]{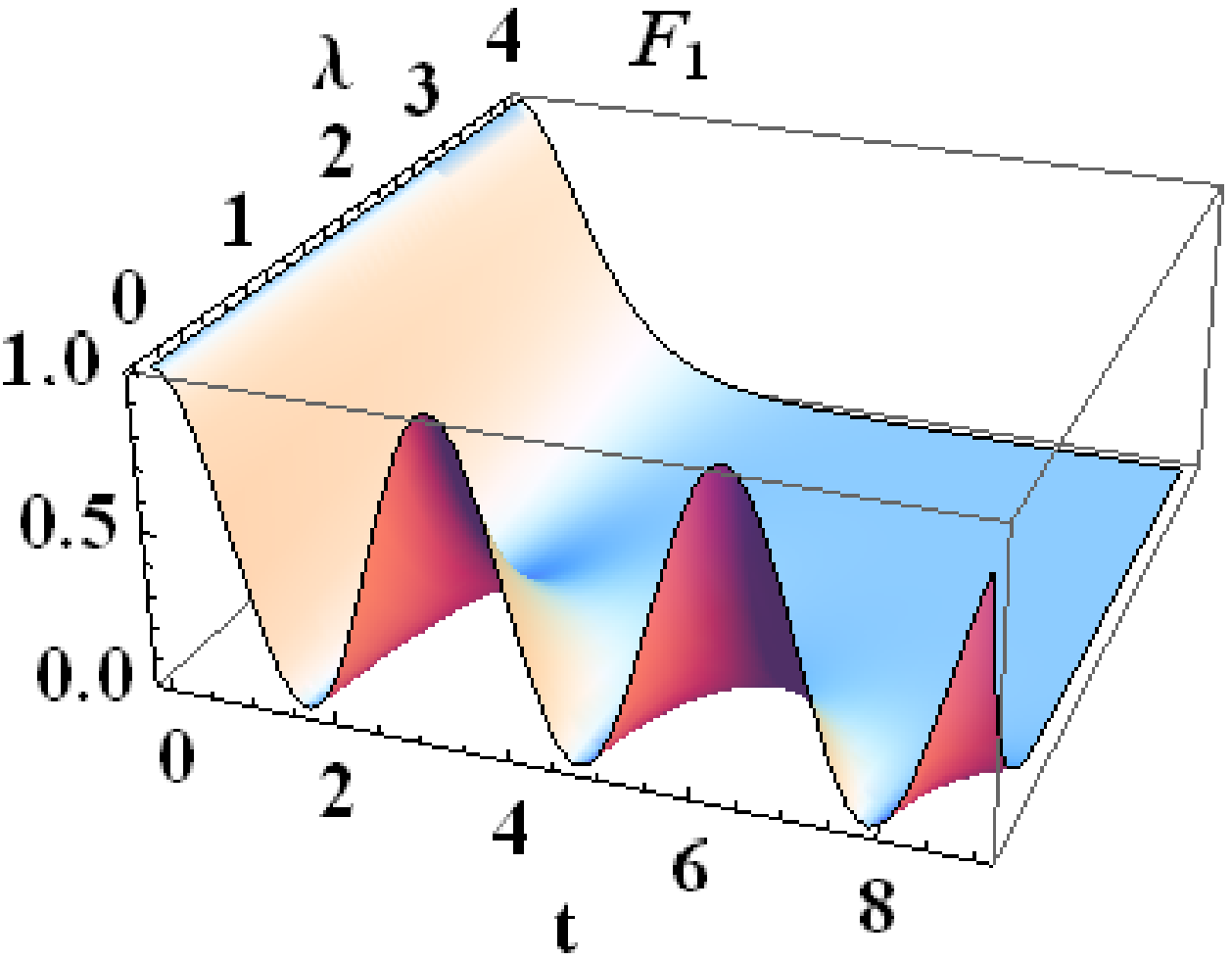}}\vspace{-1.1mm} \hspace{1.1mm}
     \subfigure{\label{fig1b}\includegraphics[width=4.5cm]{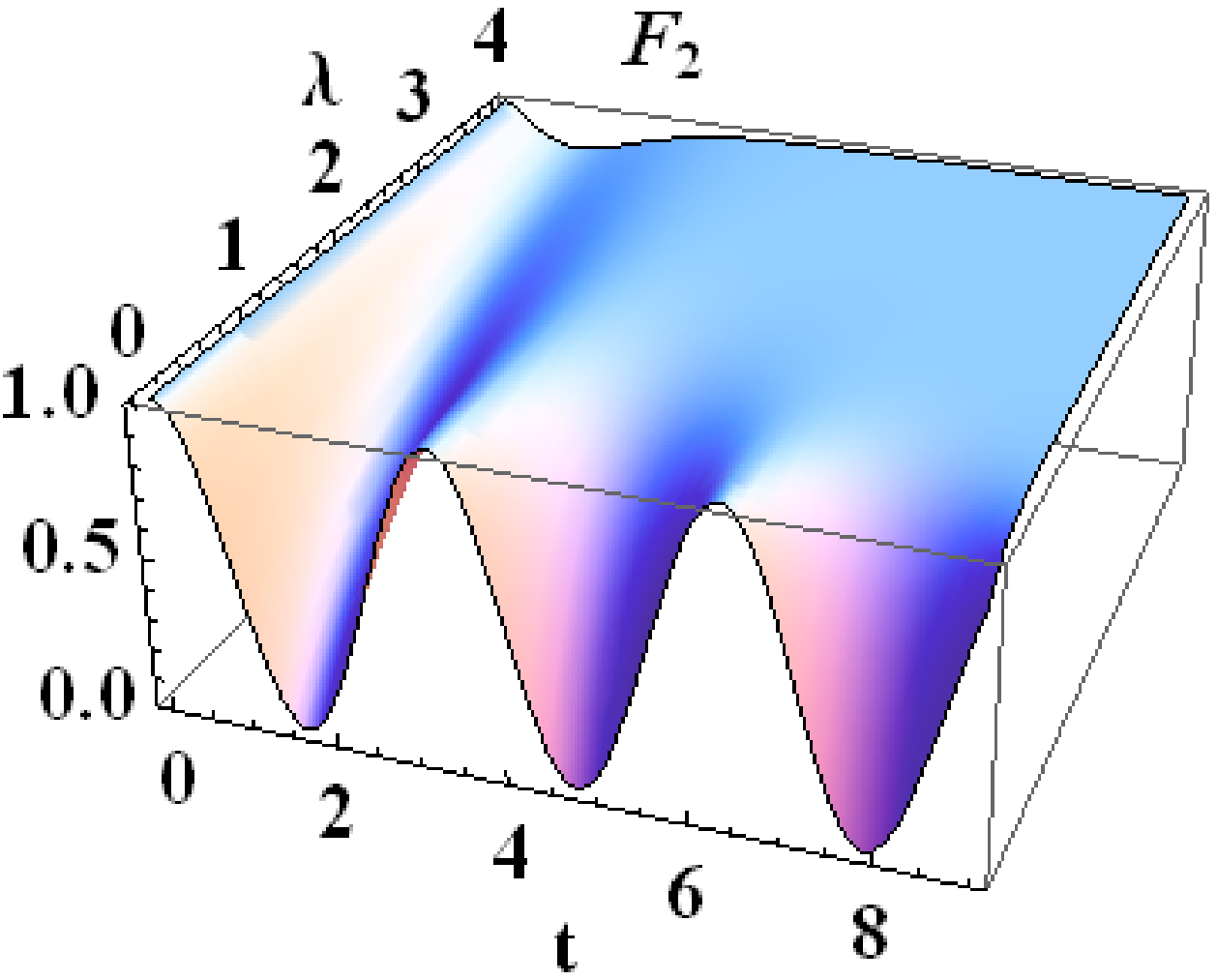}}\vspace{-1.1mm} \hspace{1.1mm}
 \subfigure{\label{fig1c}\includegraphics[width=4.5cm]{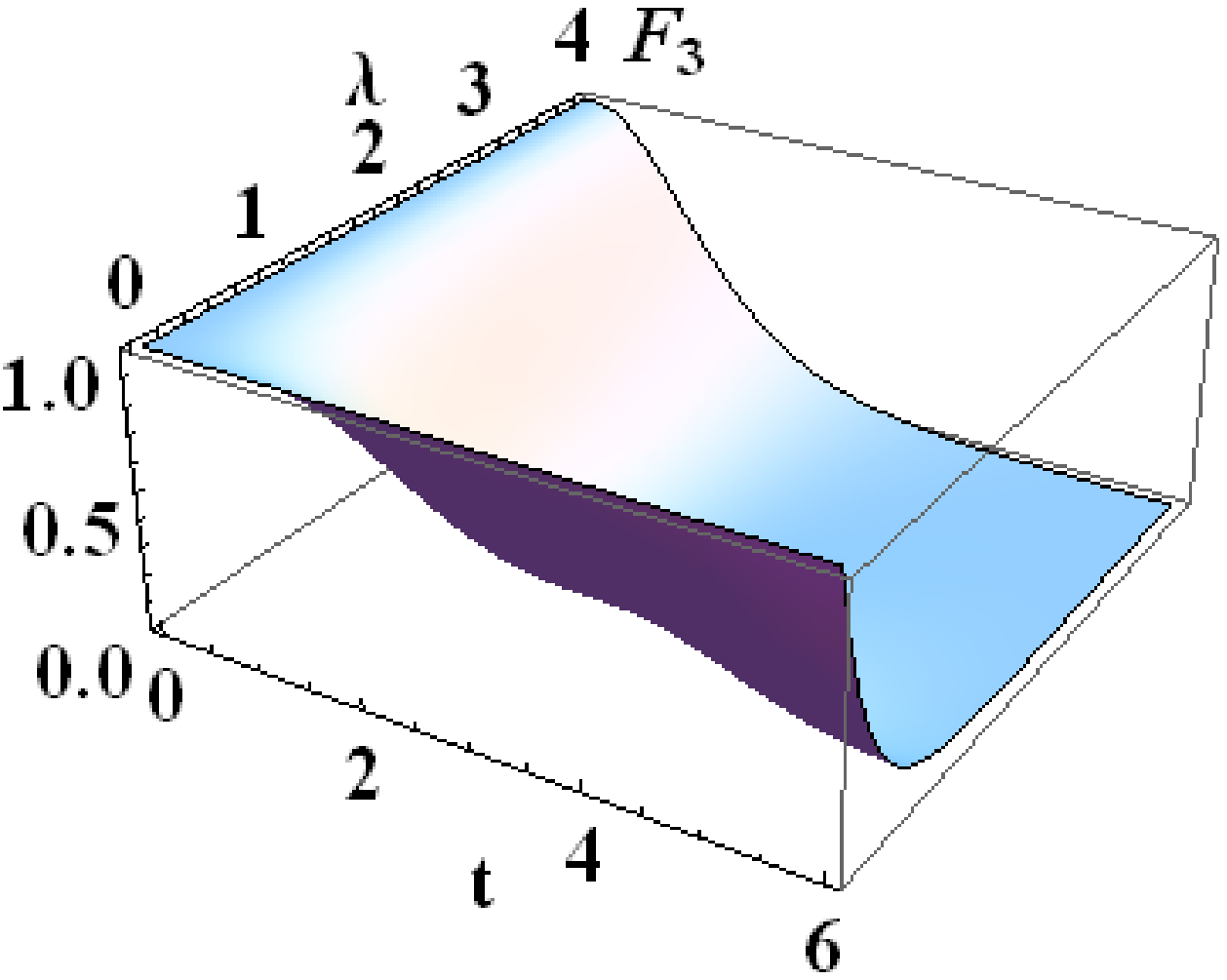}}\vspace{-1.1mm} \hspace{1.1mm}
 \subfigure{\label{fig1a}\includegraphics[width=4.5cm]{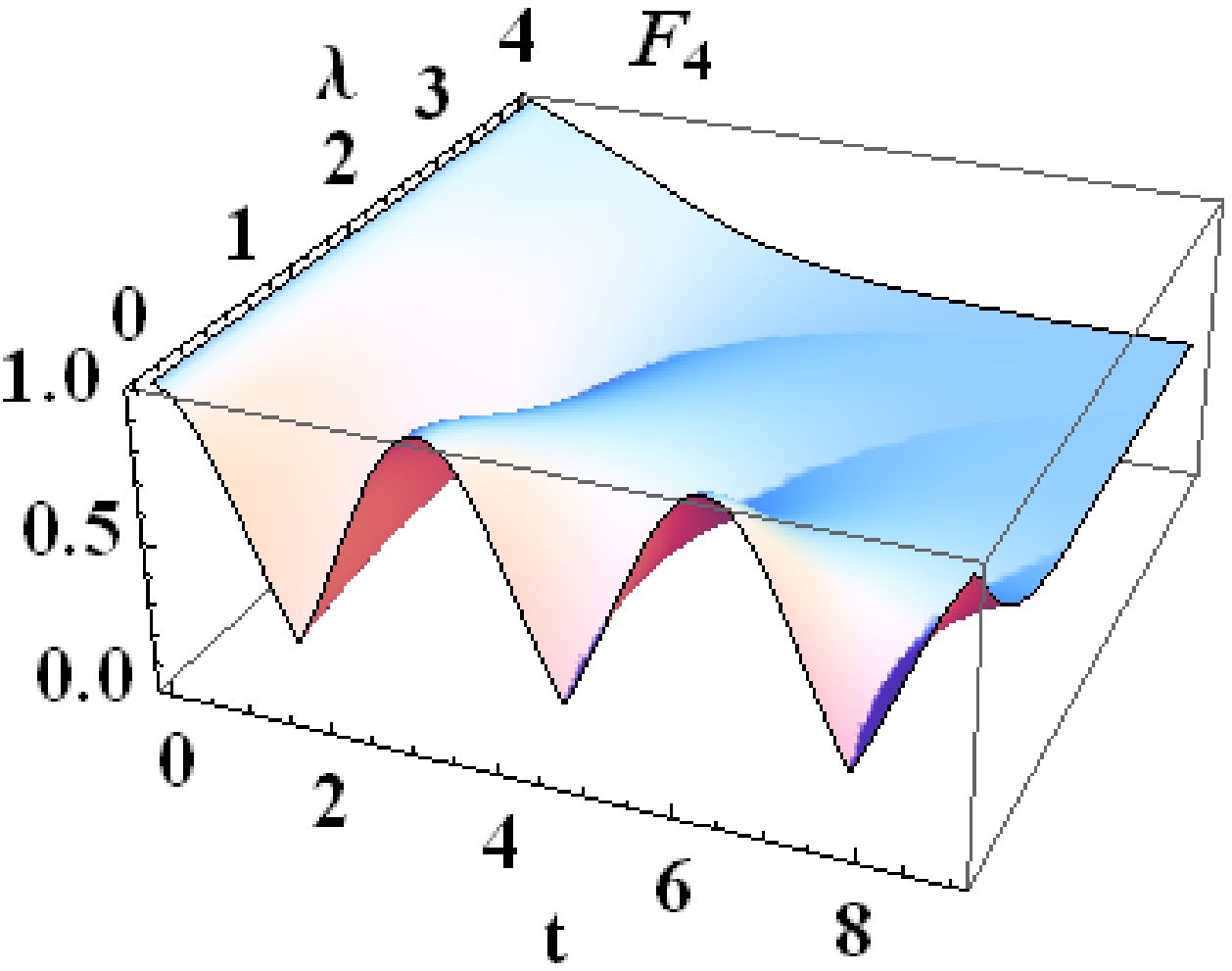}}\vspace{-1.1mm} \hspace{1.1mm}
     \subfigure{\label{fig1b}\includegraphics[width=4.5cm]{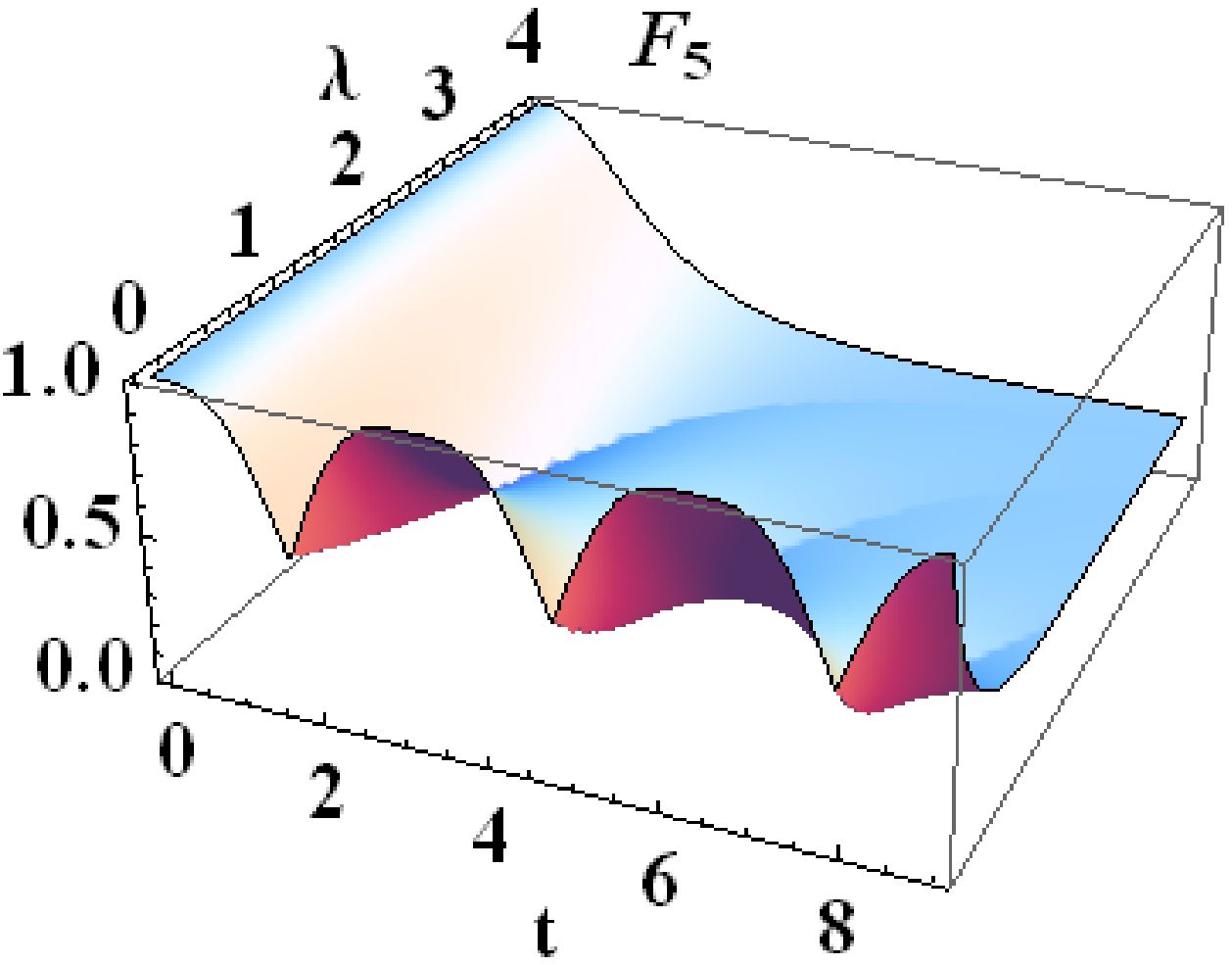}}\vspace{-1.1mm} \hspace{1.1mm}
 \subfigure{\label{fig1c}\includegraphics[width=4.5cm]{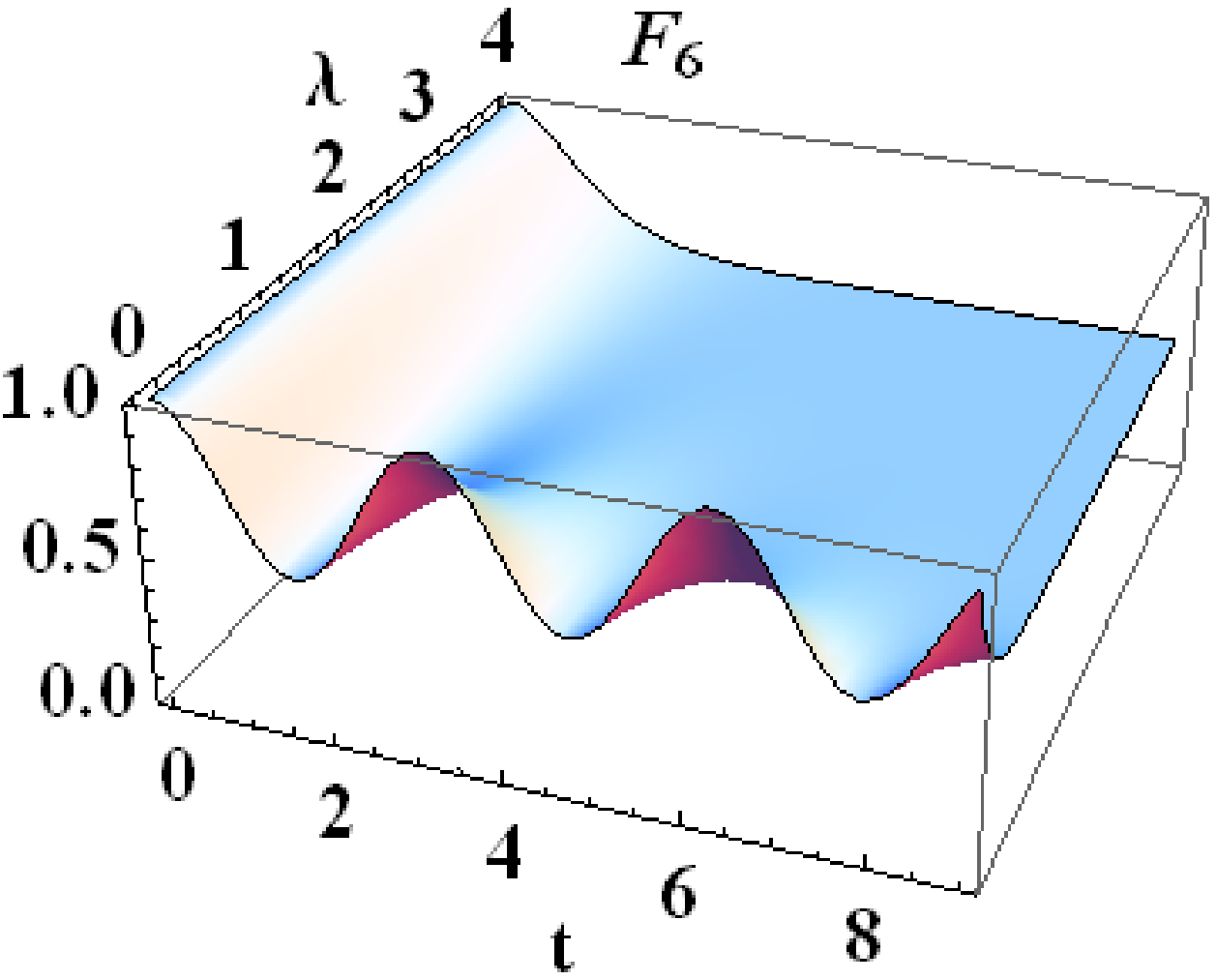}}\vspace{-1.1mm} \hspace{1.1mm}
     \end{center}
  \caption{  Fidelities $\mathit{F}_i$ (i =1..6) between various subsystem partitions as a function of
dimensionless time $t$ and decay rate $\lambda_c$. $\mathit{F}_1 \equiv$ atom-atom partition,
$\mathit{F}_2 \equiv$ cavity-cavity partition, $\mathit{F}_3 \equiv$ reservoir-reservoir  partition,
$\mathit{F}_4 \equiv$ atom-cavity partition, $\mathit{F}_5 \equiv$  atom-reservoir partition,
$\mathit{F}_6 \equiv$ cavity-reservoir partition.  The unit system adopted here and 
in all other figures is based on $\hbar=V=$1, with time $t$  obtained as inverse of 
 $\Omega_0$ (at $\lambda_c$ = 0)}
 \label{fideA}
\end{figure}

The fidelities $\mathit{F}_i$ (i=1 to 6) between various subsystems for $\lambda_c < $4, 
are shown in  Fig.~\ref{fideA}.
At the exceptional point  (setting  $\hbar=V=$1), 
$\lambda_c $=4, and minimum passage time $t$=1, $p$=1.
We note that the fidelity of the cavity-cavity partition experiences 
a minimum at $t$=1, which  appears to be a 
 characteristic feature of the cavity-cavity subsystem acting as a quantum channel.
The local  minimum in the fidelity measure  may be linked to the 
merging of two eigenvalues  at the exceptional point  \cite{Heiss} and a corresponding 
loss of distinguishability. As expected,  fidelities of all other subsystem decrease with time, and in particular
the loss in fidelities of the atom-atom  and reservoir-reservoir subsystems
are almost similar. The decrease in fidelity
of the reservoir-reservoir subsystem without any revival, is consistent with the role
 of a dissipative  sink as there is a one-directional information flow between
subspaces of ${\cal P}$  and  ${\cal Q}$, and the dissipative measure, 
$\gamma_d$ increases with time. This  feature is expected to be present in photosynthetic sinks  as well.

\begin{figure}[htp]
     \subfigure{\label{fig2a}\includegraphics[width=5.5cm]{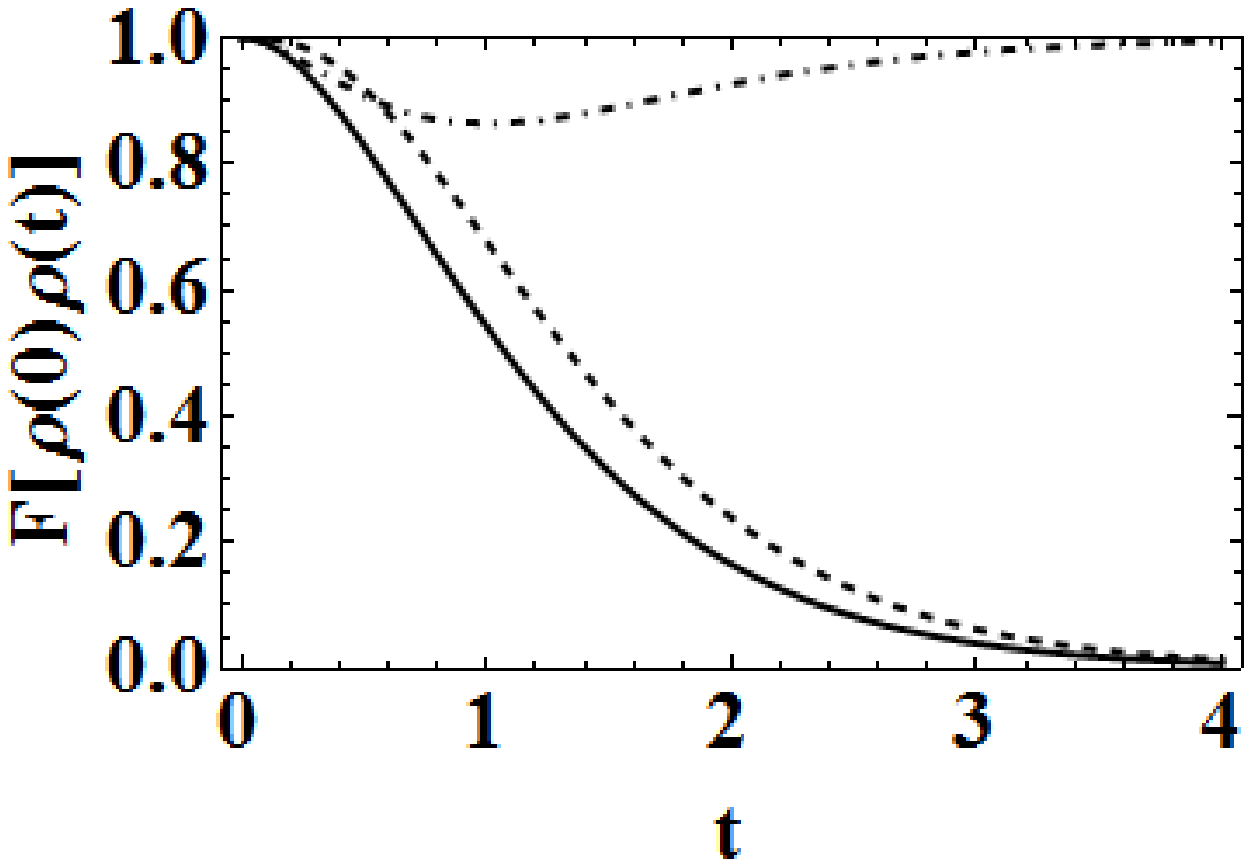}}\vspace{-1.1mm} \hspace{1.1mm}
     \subfigure{\label{fig2b}\includegraphics[width=5.5cm]{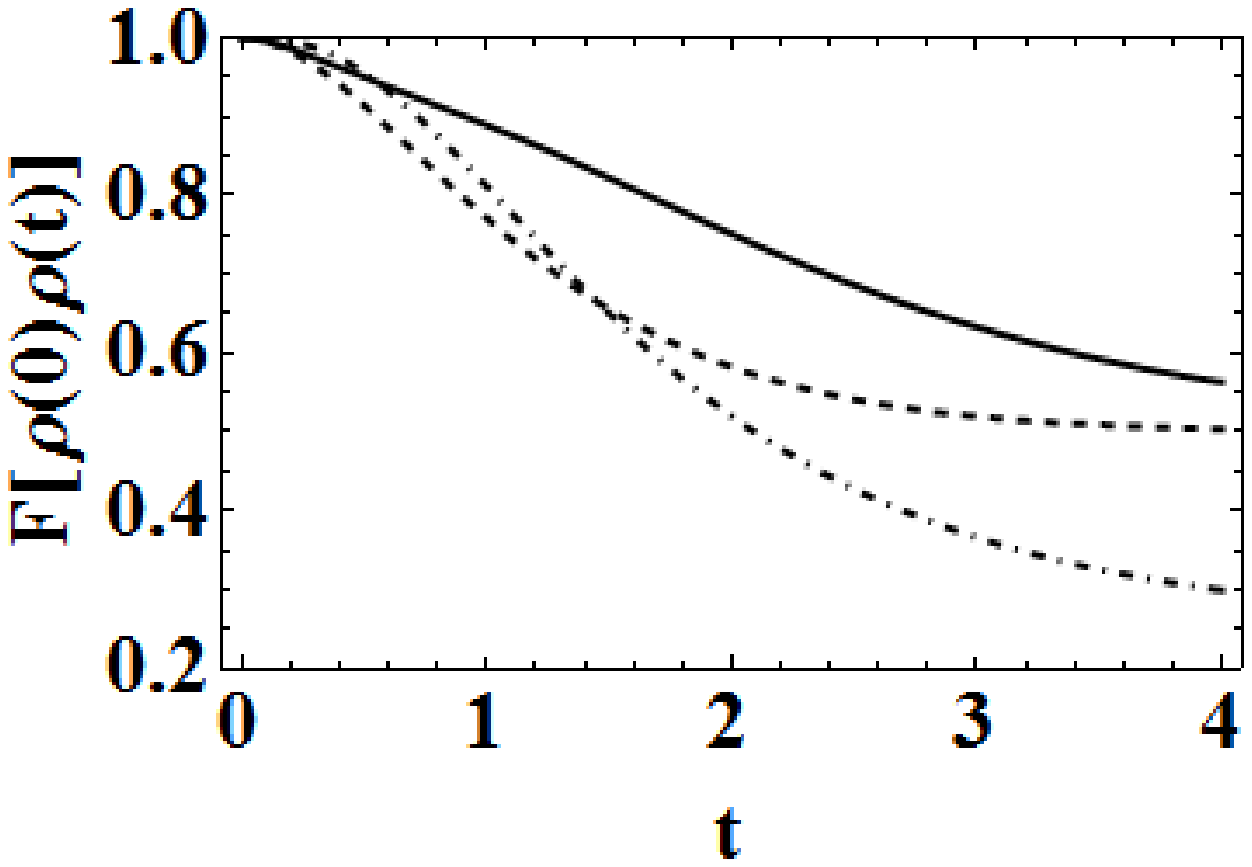}}\vspace{-1.1mm} \hspace{1.1mm}
  \caption{ Fidelities $\mathit{F}_i$ (i =1..6) of various subsystem partitions as a function of
dimensionless time $t$  at the vicinity of the exceptional point.
Solid lines ($\mathit{F}_1$, $\mathit{F}_4$), DotDashed ($\mathit{F}_2$, $\mathit{F}_5$),
Dashed ($\mathit{F}_3$, $\mathit{F}_6$) 
 }
 \label{fideB}
\end{figure}

\section{ Non-Markovianity based on the Fidelity measure}\label{fidemar}

For a system density matrix undergoing  Markovian evolution
under completely positive, trace preserving dynamical maps $\Phi$,  the fidelity measure
 $F[\rho(t),\rho(t+\tau)]$    between 
 the initial state $\rho(t)$ and the evolved state  $\rho(t+\tau)$ at a later time $t+\tau$. 
 satisfies \cite{raja} 
\bea
\label{inequa}
F[\rho(t),\rho(t+\tau)]\equiv F[\Phi(t)\rho(0),\Phi(t)\rho(\tau)] \nonumber \\ 
\Rightarrow F[\rho(t),\rho(t+\tau)]\geq F[\rho(0),\rho(\tau)].
\eea
The  violation of the inequality in Eq.~(\ref{inequa}) was proposed as
a signature of non-Markovian dynamics by Rajagopal et. al. \cite{raja}.
Accordingly negative values of the following fidelity difference function 
\be
\label{fd} 
G(t, \tau) = \frac{F[\rho(t),\rho(t+\tau)] - F[\rho(0),\rho(\tau)]}{F[\rho(0),\rho(\tau)]},
\ee
are correlated with  non-Markovianity during the  evolution dynamics
 of the quantum state. Using the two-qubit matrices which appear 
in  Eqs.(\ref{tqubitm}),  (\ref{tresm2}),  (\ref{tresm2}),
 (\ref{tresm3}) and the expressions for $p$ and $q$ in 
 Eqs.(\ref{co1}),  (\ref{co2}),  the fidelity difference function 
$G(t, \tau)$ can be evaluated as analytical expressions
of $t, \tau$ and $\lambda_c$. Due to the lengthy expressions
for these functions, we only provide numerical results which show
subtle changes of $G(t, \tau)$ with $\lambda_c$ for the various
subsystems.

\begin{figure}[htp]
  \begin{center}
    \subfigure{\label{fig3a}\includegraphics[width=3.25cm]{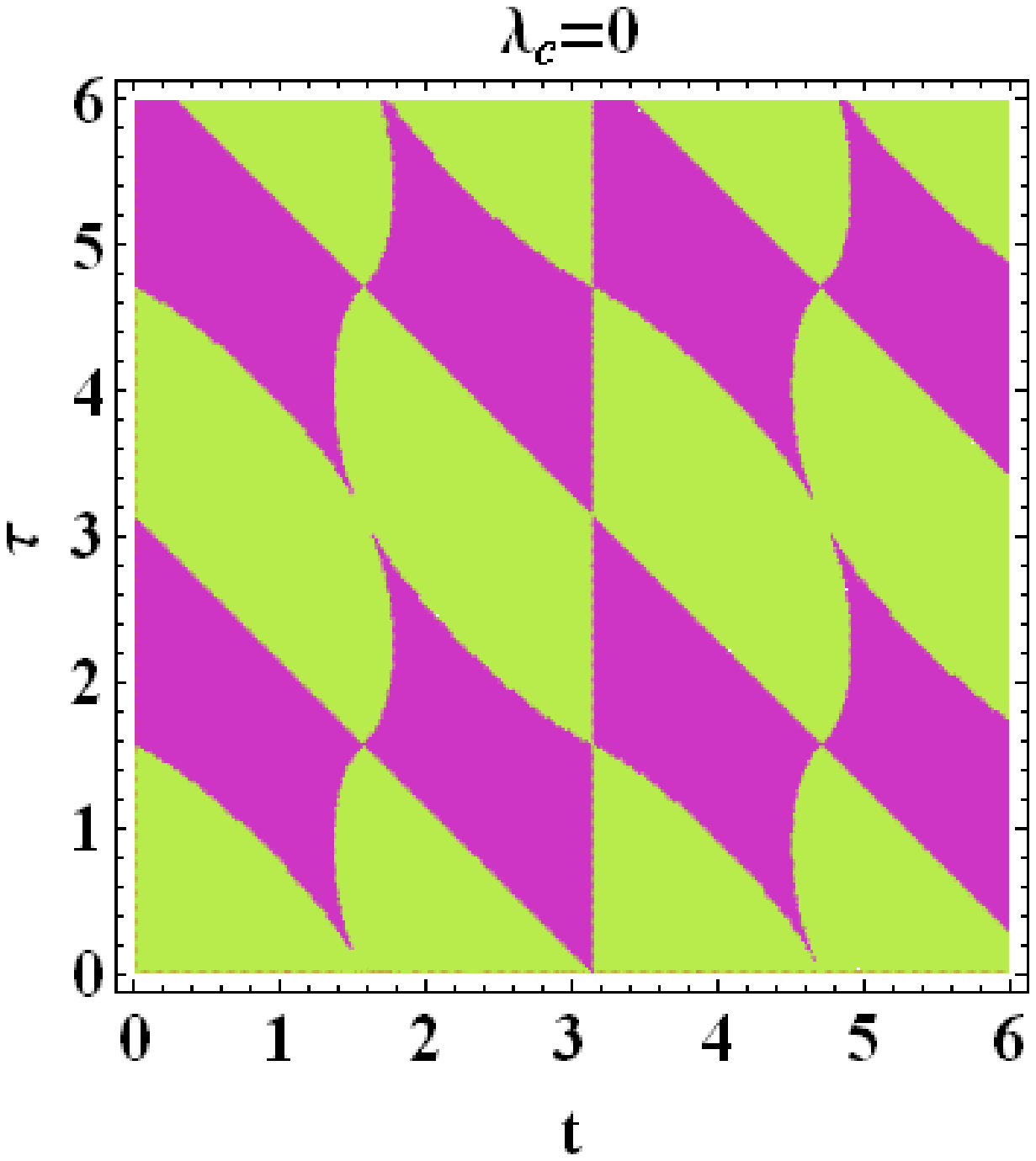}}\vspace{-1.1mm} \hspace{1.1mm}
     \subfigure{\label{fig3b}\includegraphics[width=3.25cm]{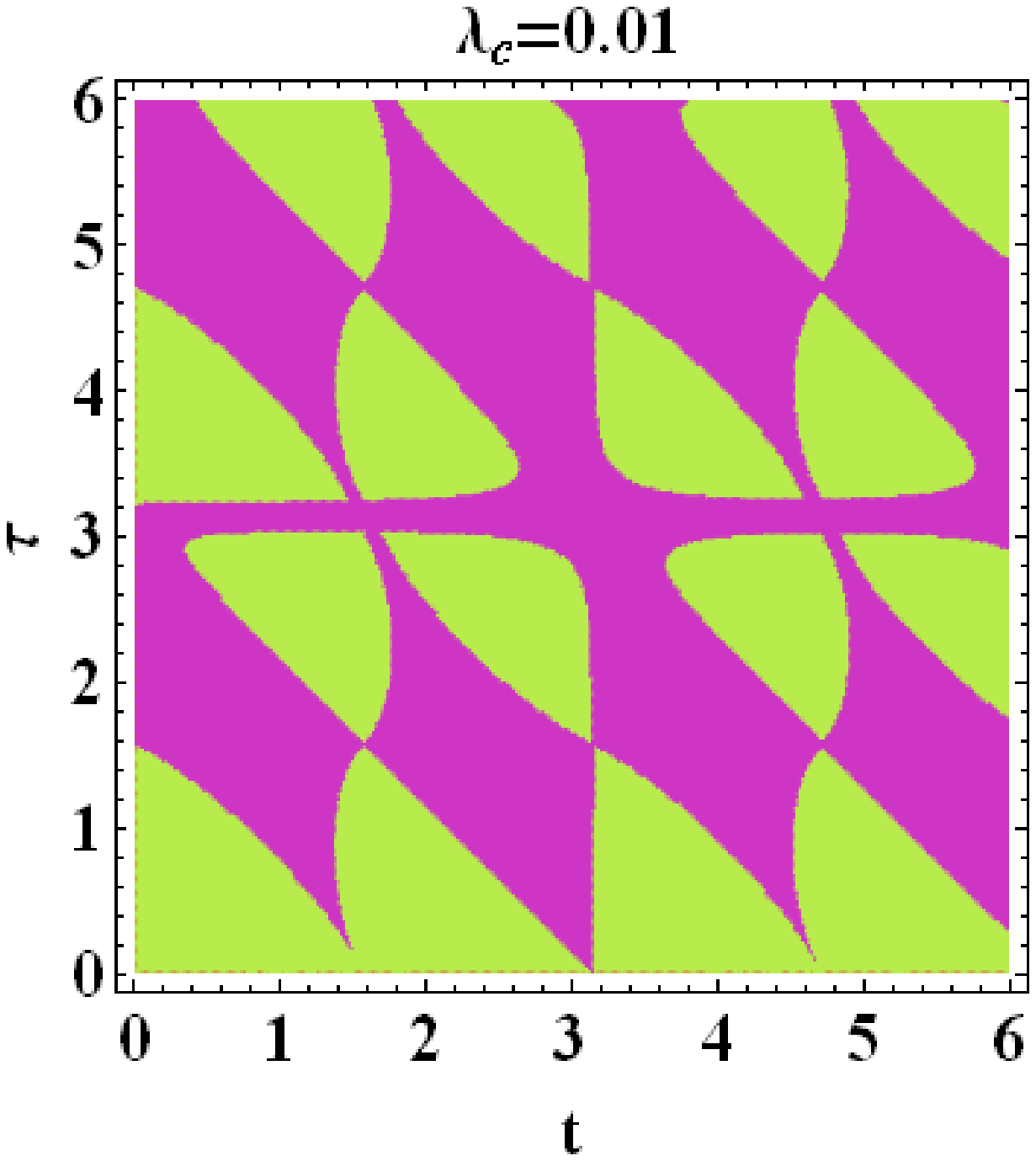}}\vspace{-1.1mm} \hspace{1.1mm}
 \subfigure{\label{fig3c}\includegraphics[width=3.25cm]{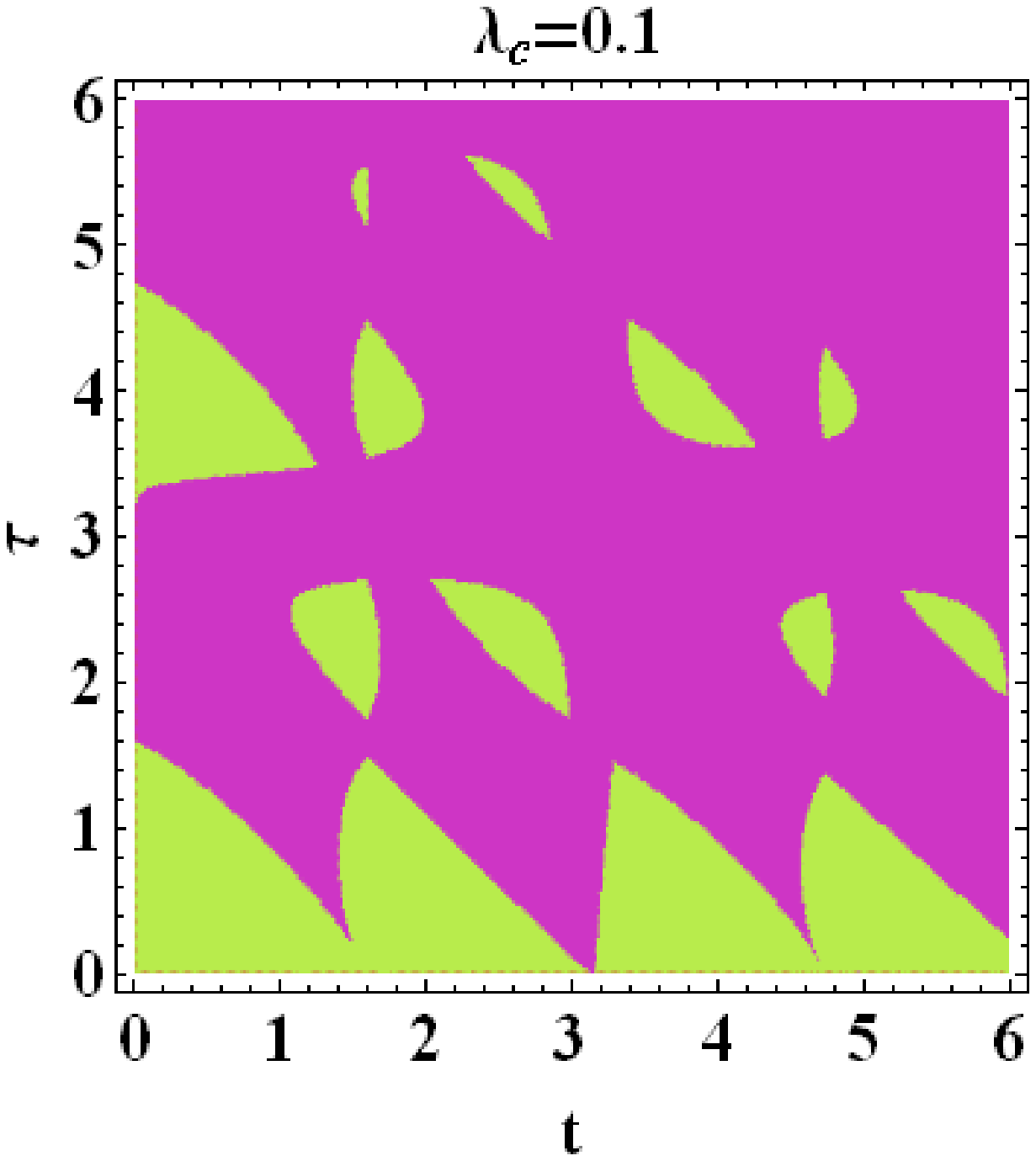}}\vspace{-1.1mm} \hspace{1.1mm}
 \subfigure{\label{fig3d}\includegraphics[width=3.25cm]{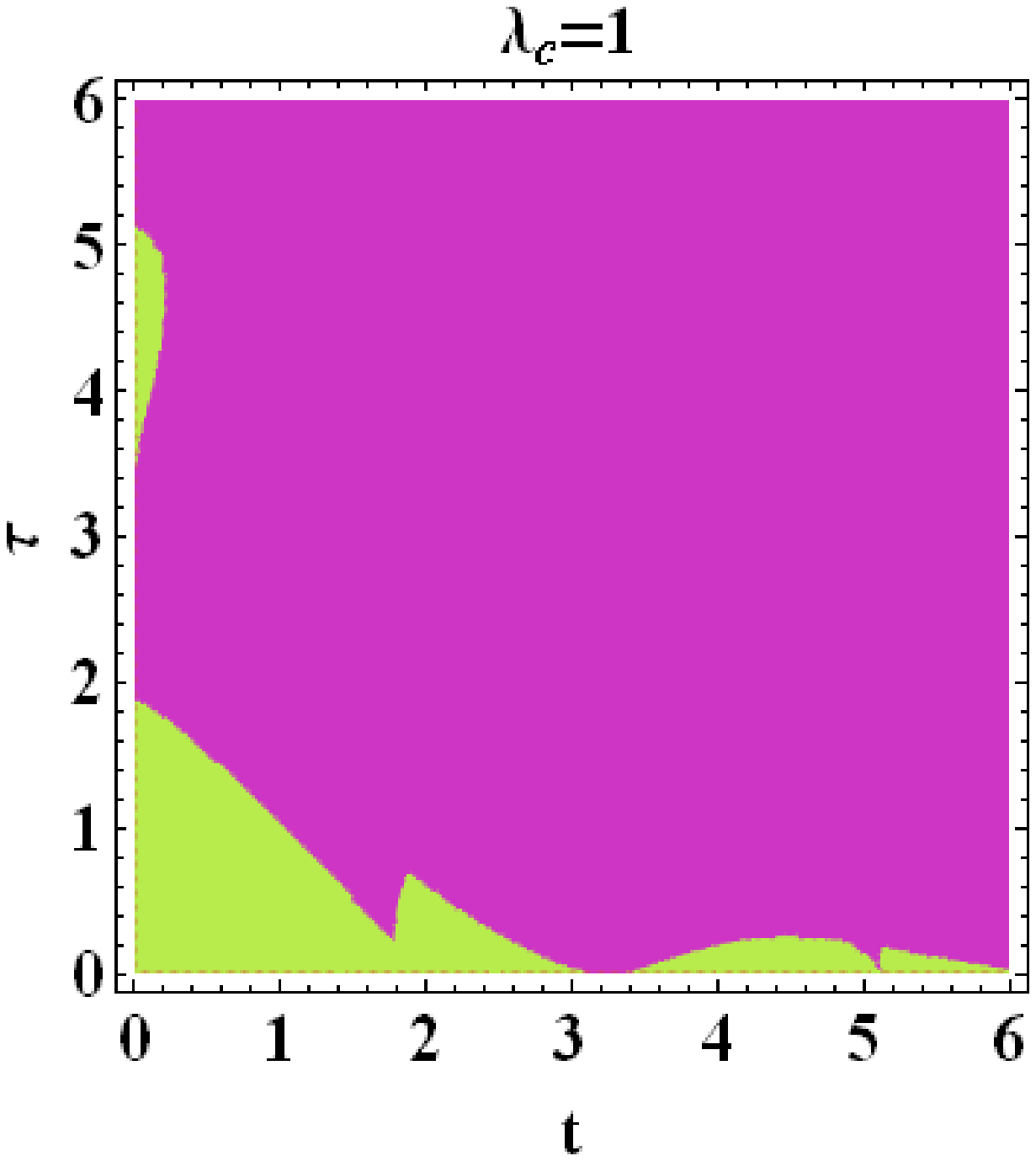}}\vspace{-1.1mm} \hspace{1.1mm}
\subfigure{\label{fig3e}\includegraphics[width=3.25cm]{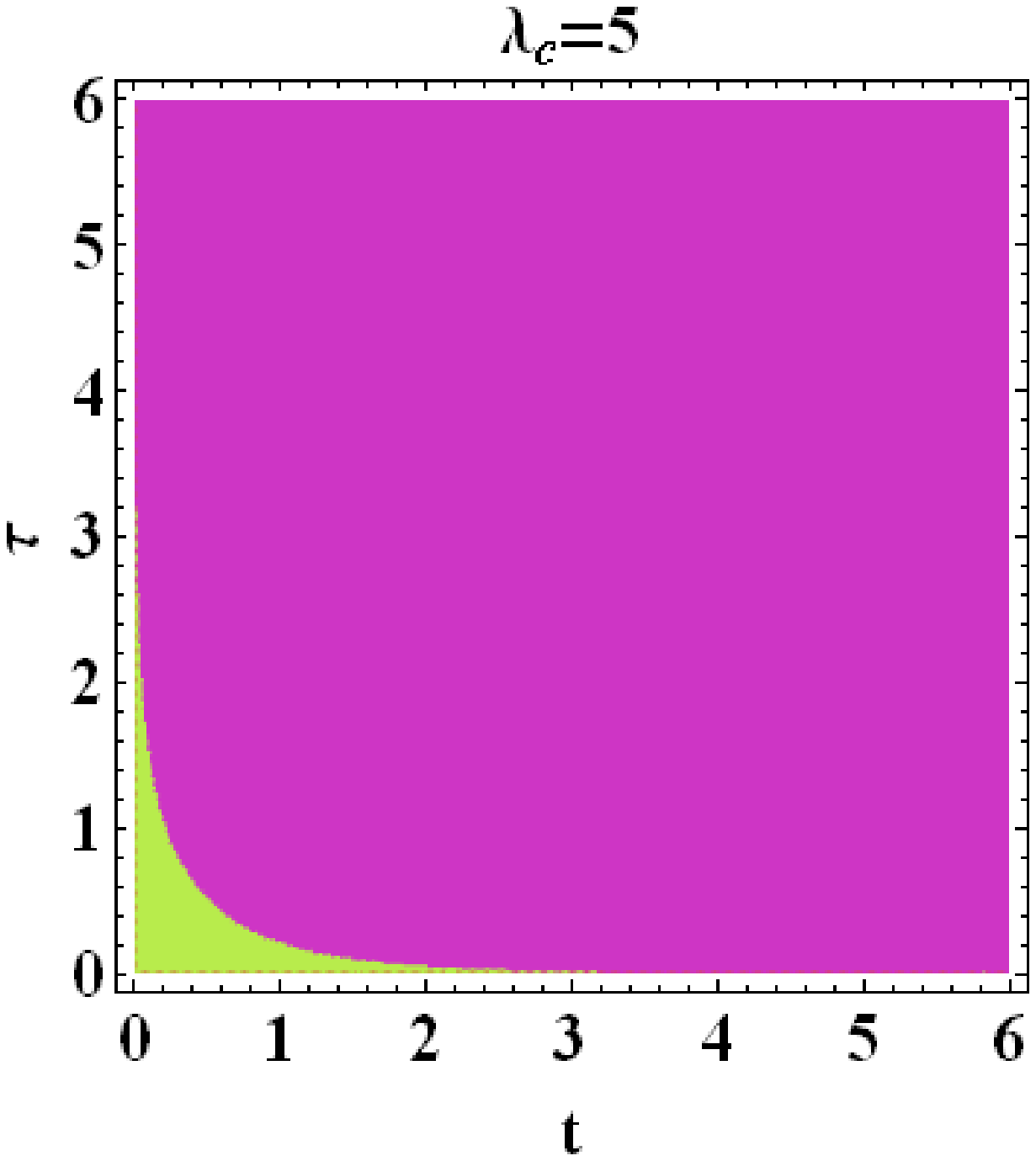}}\vspace{-1.1mm} \hspace{1.1mm}
        \end{center}
  \caption{ Fidelity difference $G(t,\tau)$  as a function of $t$ and $\tau$
for the reduced density matrix corresponding to the atom-reservoir subsystem (Eq.(\ref{tresm3})),
and  increasing $\lambda_c$. $a$=$b$=$\frac{1}{\sqrt 2}$. Non-Markovian trends were not significant
for the intra-system atom-cavity and cavity-reservoir partitions.
 Negative values   indicating non-Markovianity are shaded green, while regions which are
shaded purple imply non-violation of the inequality (\ref{inequa}) and hence Markovian evolution dynamics.
Non-Markovianity decreases with higher values of the cavity decay rate $\lambda_c$.
 }
 \label{fideC}
\end{figure}

Fig.~\ref{fideC} illustrates
the fidelity difference $G(t,\tau)$  as a function of $t$ and $\tau$
for the reduced density matrix corresponding to the atom-reservoir subsystem (Eq.(\ref{tresm3}))
for increasing $\lambda_c$. 
The sensitivity of the two-qubit dynamics on $\tau$
translates in a striking way to the
non-Markovianity measure, with  the qubit dynamics becoming
Markovian beyond a critical damping $\lambda_c$, and as expected depends
on the initial correlation  amplitude factors, $a,b$.
At $\lambda_c$=0,  the oscillatory state of  information  exchanges between 
the atomic qubit and reservoir 
 is captured by the symmetric pattern of non-Markovian dynamics alternating with 
Markovian dynamics. The non-Markovianity 
decreases with higher values of the cavity decay rate $\lambda_c$, consistent with 
decreased population transfer between the atom-cavity subsystem.
Specifically,  we note the persistence of  non-Markovianity at small $t,\tau$ values,
even at higher values of the cavity decay rate $\lambda_c \approx$ 5.

As expected, there is insignificant non-Markovian signatures  seen in a 
 atom-cavity partition  within the same subsystem.
In contrast to the lack of non-Markovianity in the  atom-cavity partition ,
we  noted richness in non-Markovian dynamics for the   atom-reservoir (${\rho}_{a_1,r_2}$ ($i \neq j$) 
partition across different subsystems, and similar to that    shown in  Fig.~\ref{fideC}. 
In this case, the two-qubit dynamics becomes fully
Markovian at  a smaller critical damping  if a  small initial correlation  amplitude factor, $a$
is used. 

\begin{figure}[htp]
  \begin{center}
    \subfigure{\label{fig4a}\includegraphics[width=3.25cm]{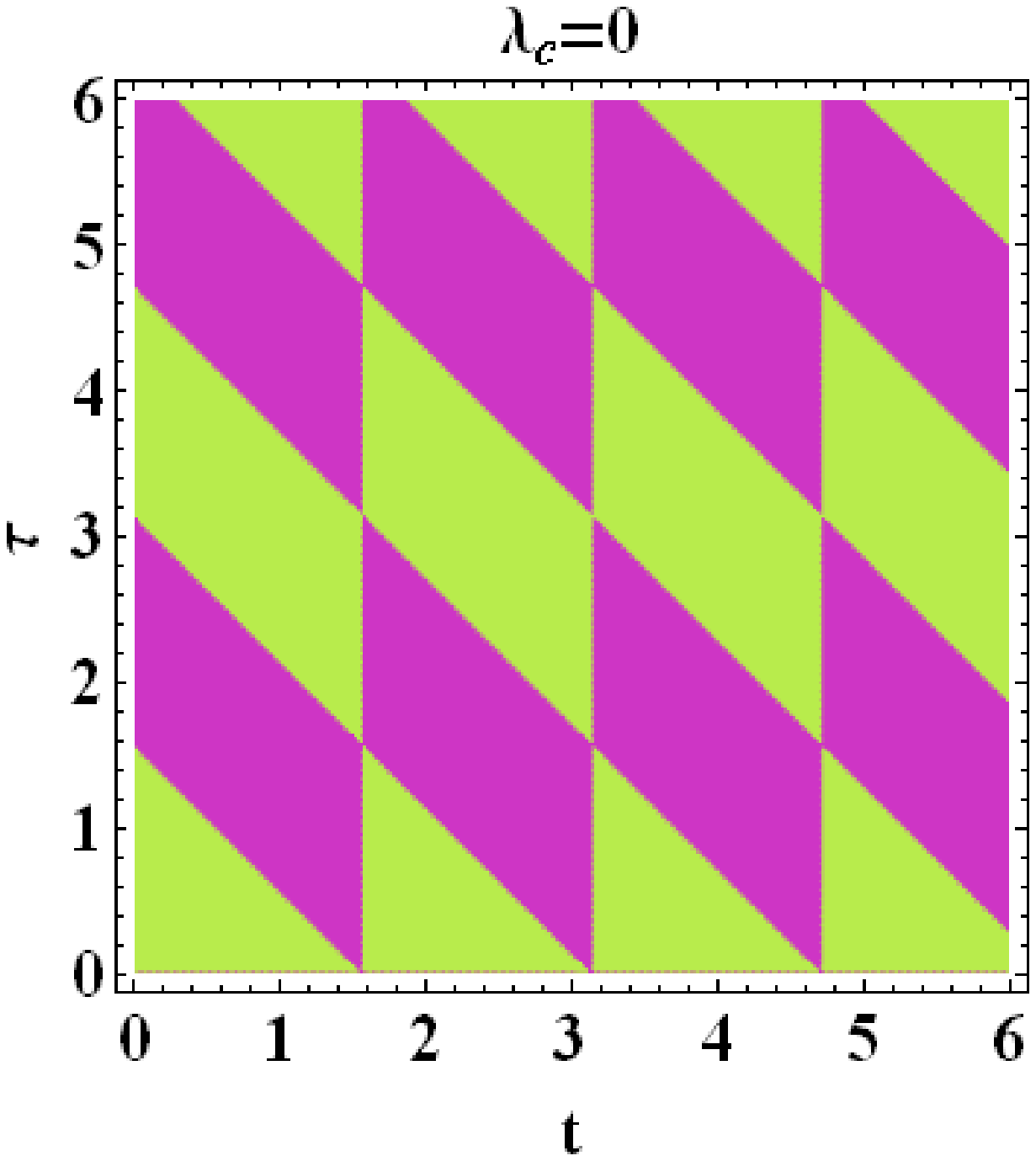}}\vspace{-1.1mm} \hspace{1.1mm}
     \subfigure{\label{fig4b}\includegraphics[width=3.25cm]{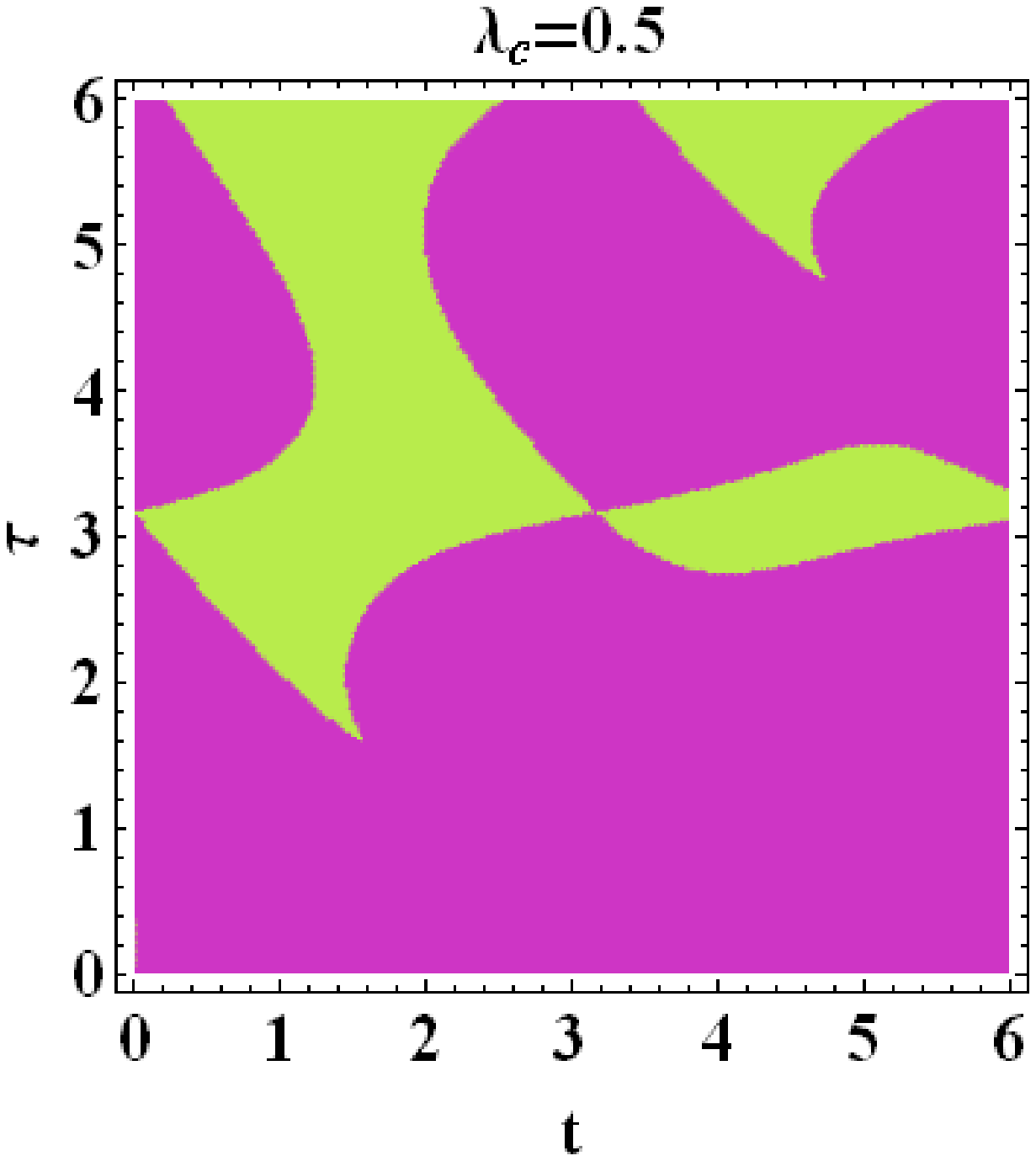}}\vspace{-1.1mm} \hspace{1.1mm}
 \subfigure{\label{fig4c}\includegraphics[width=3.25cm]{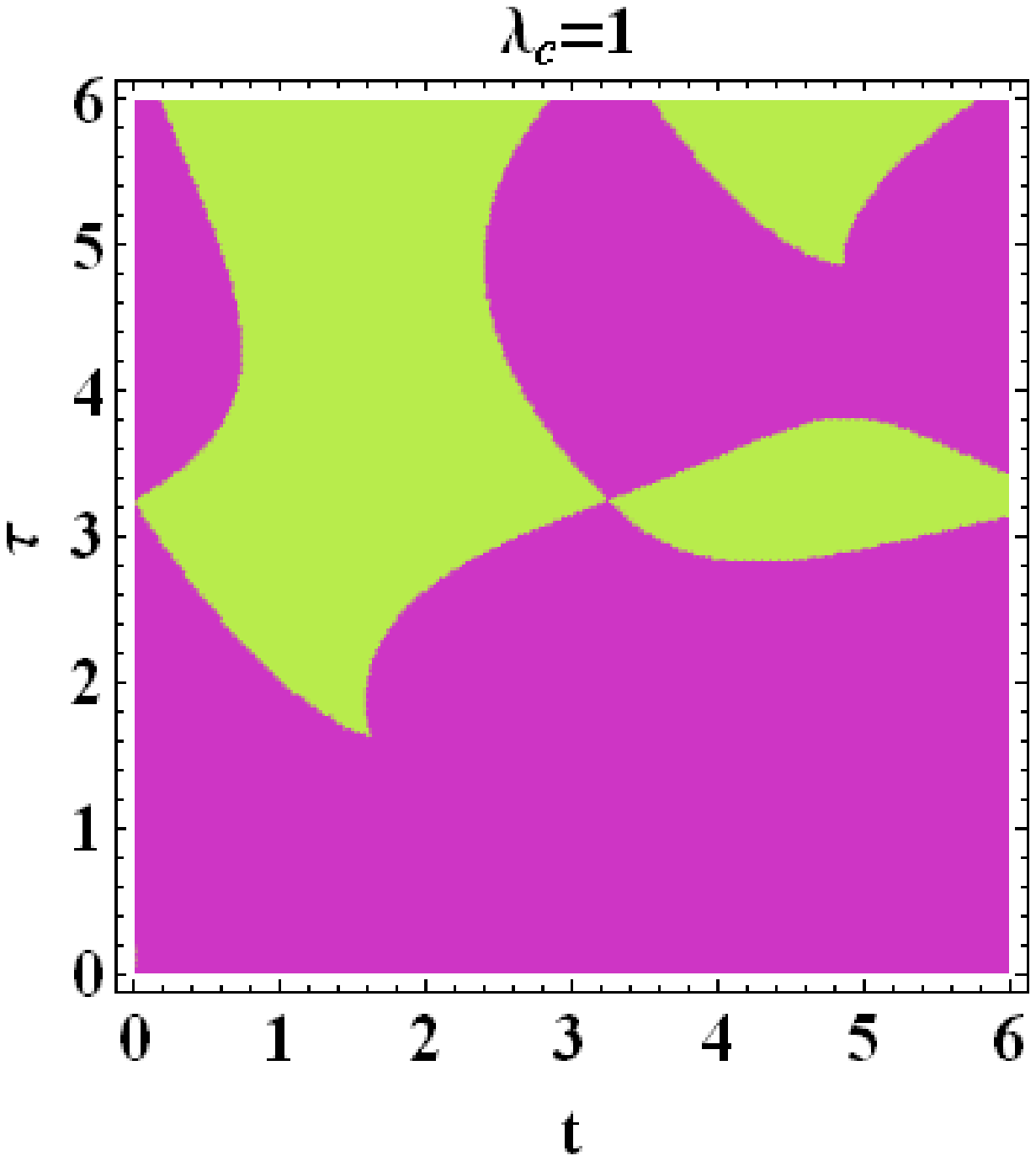}}\vspace{-1.1mm} \hspace{1.1mm}
 \subfigure{\label{fig4d}\includegraphics[width=3.25cm]{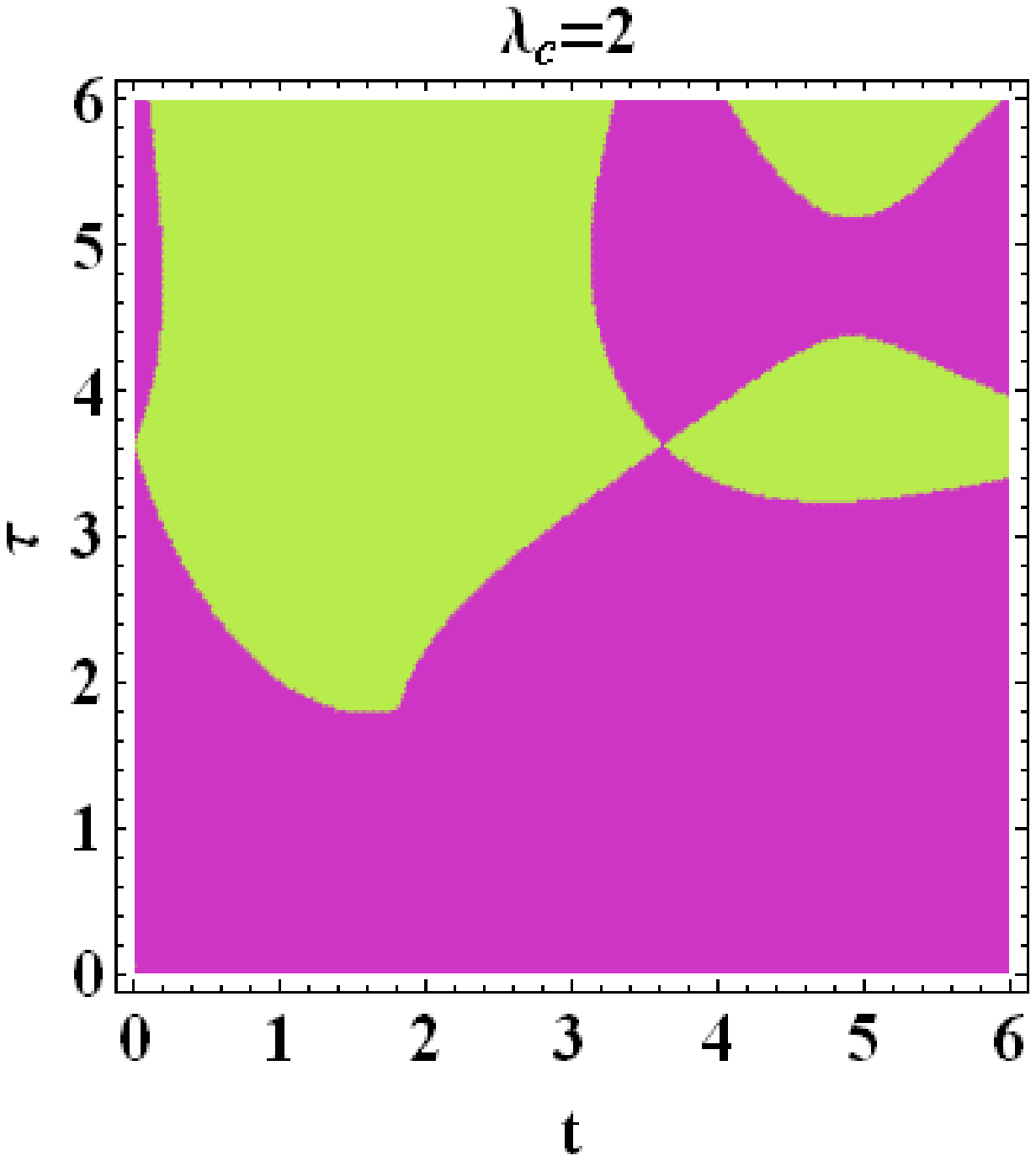}}\vspace{-1.1mm} \hspace{1.1mm}
\subfigure{\label{fig4e}\includegraphics[width=3.25cm]{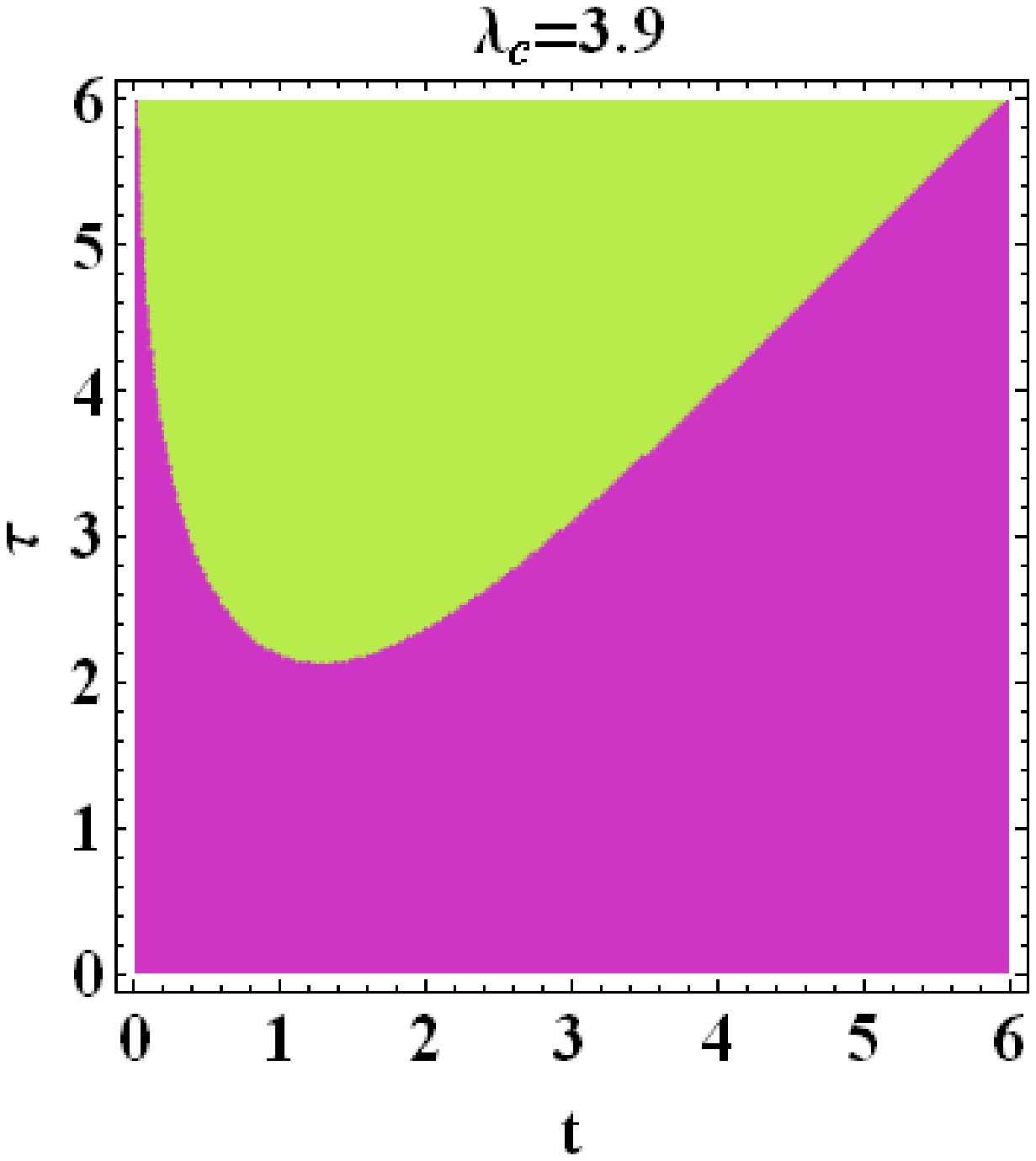}}\vspace{-1.1mm} \hspace{1.1mm}
        \end{center}
  \caption{Fidelity difference $G(t,\tau)$  as a function of $t$ and $\tau$
for the reduced density matrix corresponding to the cavity-cavity subsystem ( Eq.(\ref{tqubitm}) with
 the substitution $(1-p)\rightarrow q$). $a$=$b$=$\frac{1}{\sqrt 2}$.
 Colour codes follows as specified in Fig.~\ref{fideC}.
Unlike Fig.~\ref{fideC}, non-Markovianity continue  to persist 
for higher values of the cavity decay rate $\lambda_c$ in  some regions. 
}
 \label{fideD}
\end{figure}

The fidelity difference $G(t,\tau)$  is also seen to undergo notable changes (with $t$ and $\tau$)
in the case of the reduced density matrix corresponding to the cavity-cavity subsystem ( Eq.(\ref{tqubitm}) with
 the substitution $(1-p)\rightarrow q$), as shown in Fig.~\ref{fideD}. Here,
the   non-Markovianity appear to be enhanced in some regions as the 
cavity decay rate $\lambda_c$ is increased. In the vicinity of the exceptional point,
we note a transition in which regions of non-Markovianity merge, and Markovian processes
are eliminated.  The  exceptional point appear to signal the enhancement
of non-Markovianity in selected regions (of time $t$ and $\lambda_c$)
for the cavity-cavity partition. 

\subsection{ Non-Markovianity based on other distance measures}\label{fidemarC}

It may be worthwhile to examine other physical quantities, which are functions of the dynamical map, that describe the open system evolution of the physical states. In an earlier work \cite{usud}, 
the use of relative entropy difference  to qualitatively capture the departure from the completely positive Markovian semigroup property of evolution was verified to assume negative values under an open system  non-completely positive dynamics.
 We recall that the relative entropy~\cite{veple} of two density matrices $\rho$ and $\gamma$, defined by: 
$ S(\rho\vert\vert\gamma)={\rm Tr}[\rho(\ln\rho-\ln\gamma)]$  is positive  and  vanishes if and only if $\rho\equiv \gamma$. Under completely positive, trace preserving dynamical maps $\Phi$,  the relative entropy 
obeys monotonicity property which yields  a relation which is analogous to Eq.(\ref{inequa})
\bea
\label{srem}
S[\rho(t)\vert\vert\rho(t+\tau)]&\equiv& S[\Phi(t)\rho(0)\vert\vert \Phi(t)\rho(\tau)]\nonumber \\ 
 &\leq & S[\rho(0)\vert\vert\rho(\tau)]
\eea
 The  relative entropy difference $S(t,\tau)$ given by
\be
\label{relD}
S(t,\tau)=\frac{S[\rho(0)\vert\vert\rho(\tau)]-S[\rho(t)\vert\vert\rho(t+\tau)]}{S[\rho(0)\vert\vert\rho(\tau)]}
\ee
is necessarily positive for all quantum states $\rho(t)$ evolving under completely positive Markovian dynamics and
 violation of the inequality in Eq.(\ref{srem}) i.e.,
$S(t,\tau) < 0$, denotes the occurrence of non-Markovian  dynamical processes in the quantum system under
study.  

\begin{figure}[htp]
  \begin{center}
    \subfigure{\label{fig5a}\includegraphics[width=3.25cm]{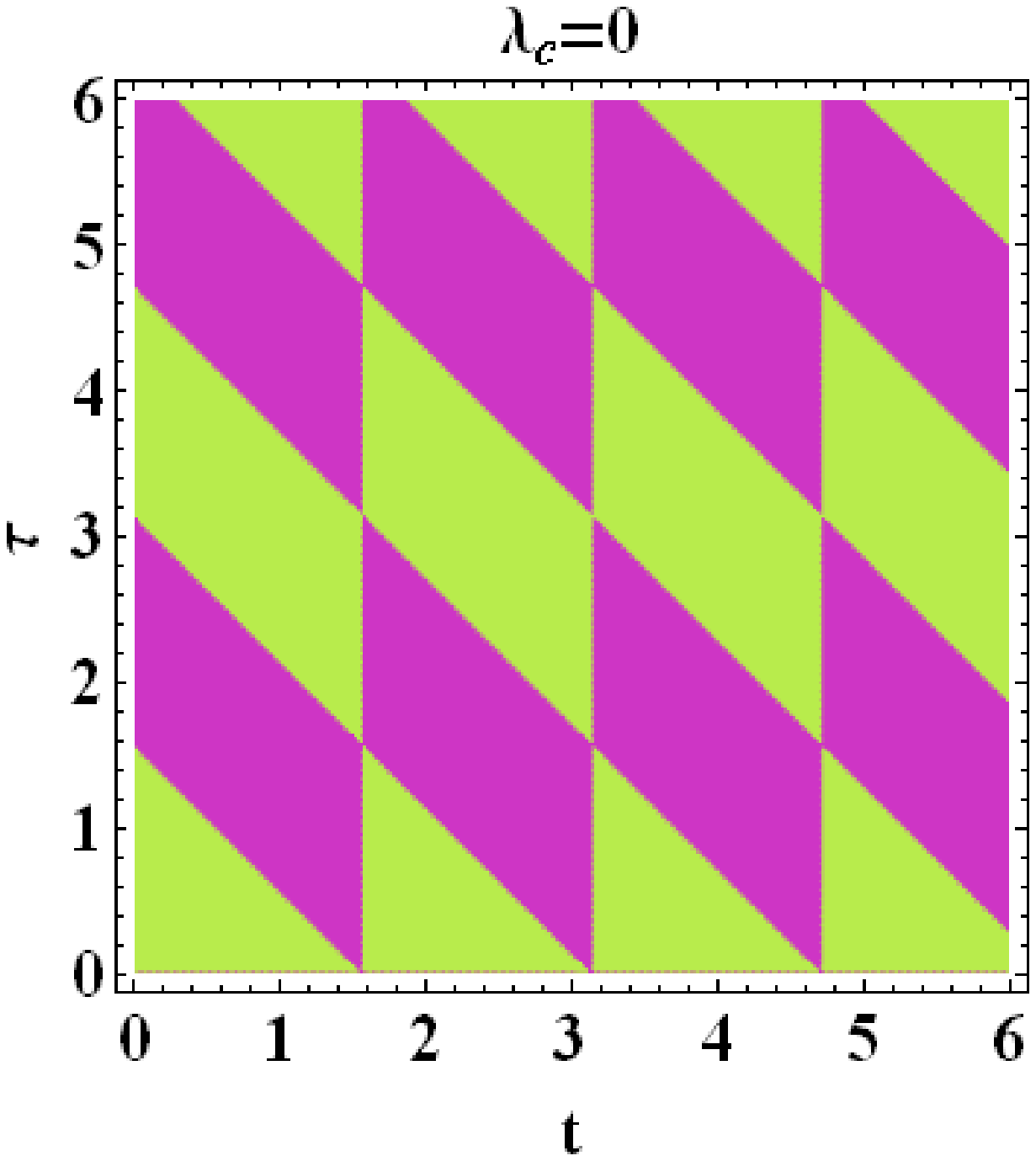}}\vspace{-1.1mm} \hspace{1.1mm}
     \subfigure{\label{fig5b}\includegraphics[width=3.25cm]{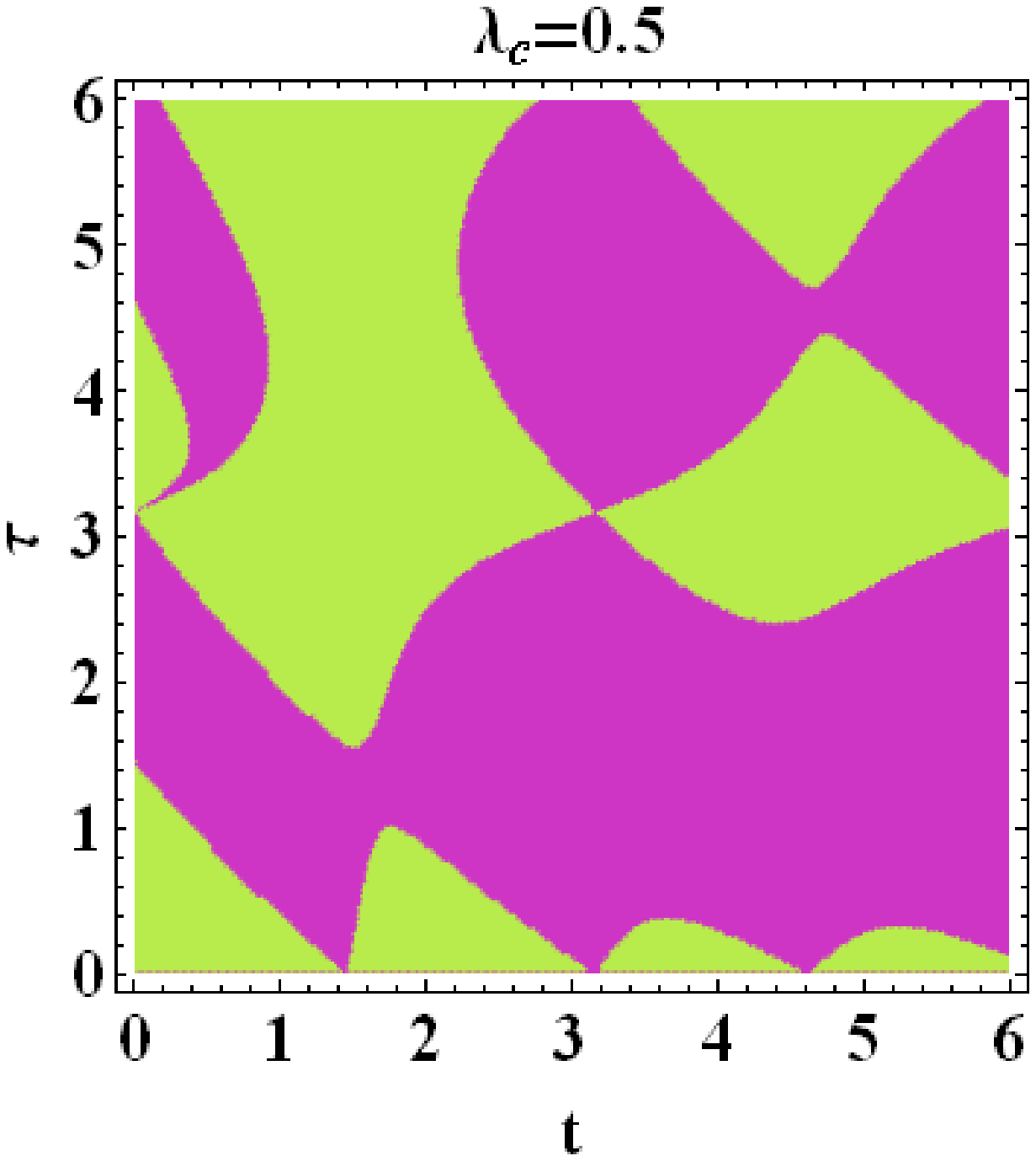}}\vspace{-1.1mm} \hspace{1.1mm}
 \subfigure{\label{fig5c}\includegraphics[width=3.25cm]{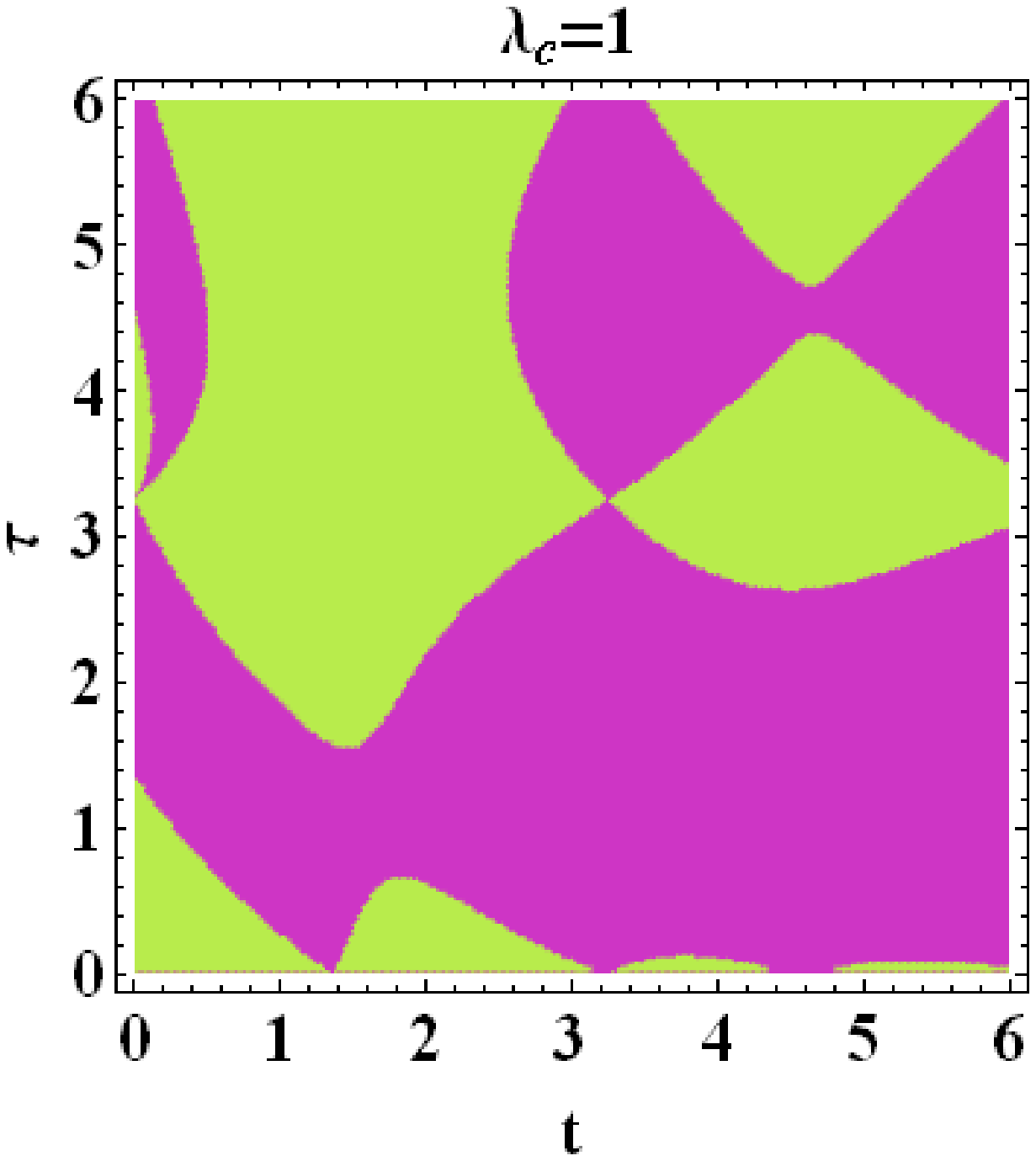}}\vspace{-1.1mm} \hspace{1.1mm}
 \subfigure{\label{fig5d}\includegraphics[width=3.25cm]{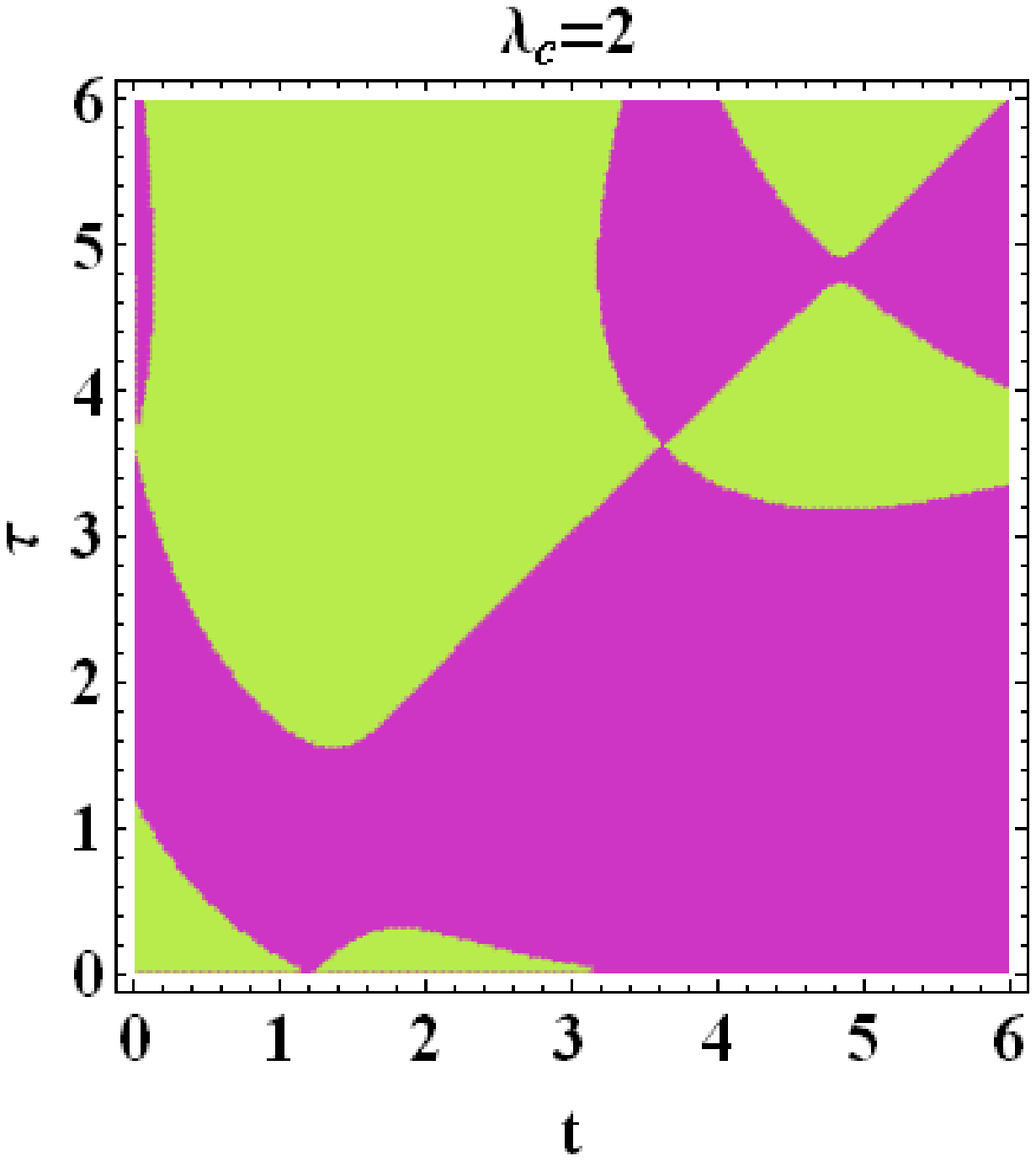}}\vspace{-1.1mm} \hspace{1.1mm}
\subfigure{\label{fig5e}\includegraphics[width=3.25cm]{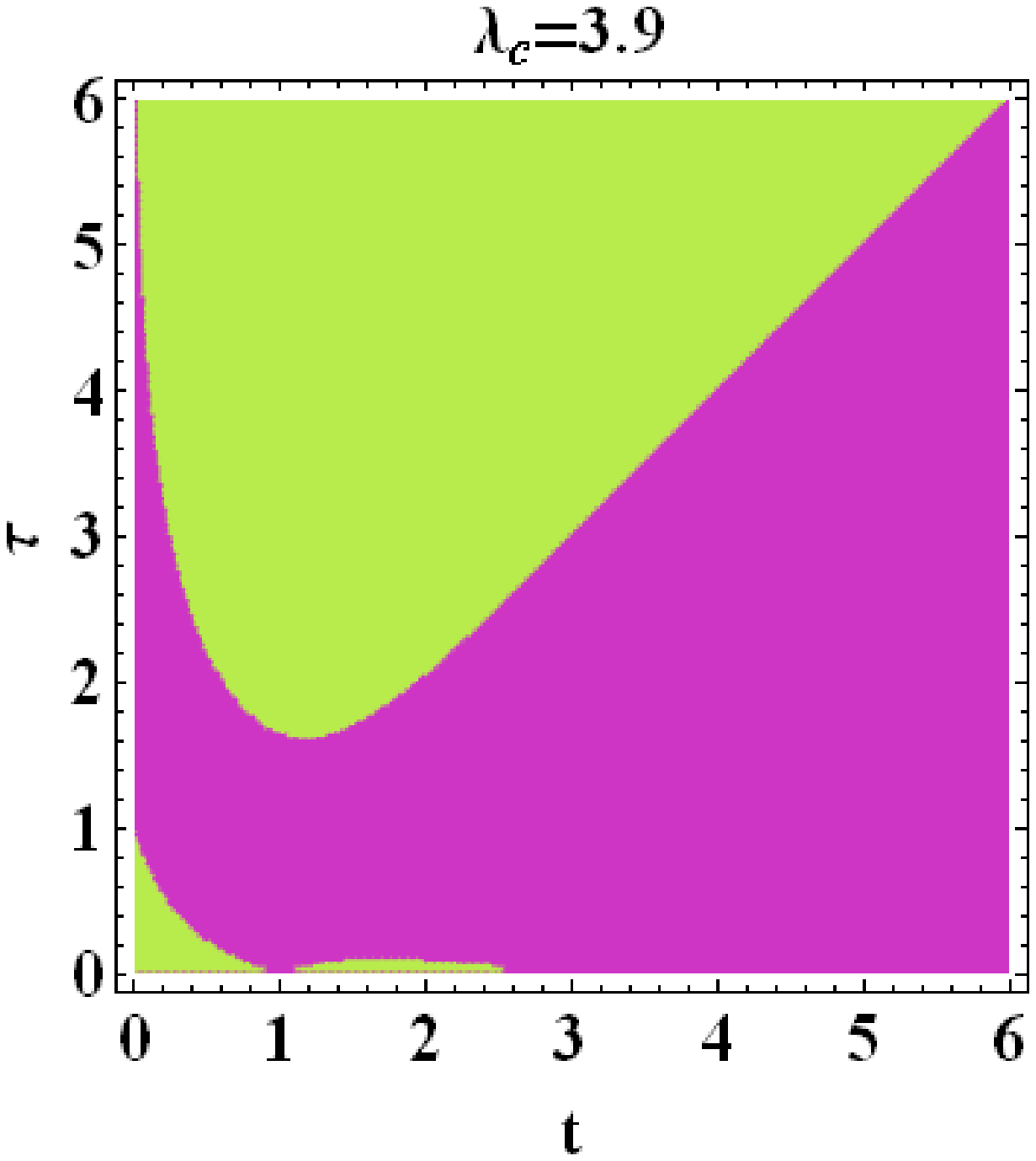}}\vspace{-1.1mm} \hspace{1.1mm}
        \end{center}
  \caption{Trace difference $D(t,\tau)$  (Eq.(\ref{traceD}) as a function of $t$ and $\tau$
for the reduced density matrix corresponding to the cavity-cavity subsystem ( Eq.(\ref{tqubitm}) with
 the substitution $(1-p)\rightarrow q$). $a$=$b$=$\frac{1}{\sqrt 2}$.
 Colour codes follows as specified in Fig.~\ref{fideC}.
Unlike Fig.~\ref{fideD}, non-Markovianity appear to be present
 at small $\tau, t$ that remains undetected by the fidelity difference measure.
In any case, there is overall agreement with Fig.~\ref{fideD} in that
non-Markovianity is optimized in the  vicinity of the exceptional point.
}
 \label{fideE}
\end{figure}

Similar to the relative entropy difference, the  trace-distance $D[\rho_1,\rho_2]=\frac{1}{2}||\rho_1-\rho_2||$ \cite{Niel}
where $||A||=Tr[\sqrt{A^\dagger A}]$  between   $\rho_{1},\ \rho_2$,
 can be used to define the trace-distance difference
\be
\label{traceD}
D(t, \tau) = \frac{D[\rho(0),\rho(\tau) - D[\rho(t),\rho(t+\tau)]}{D[\rho(0),\rho(\tau)]},
\ee
and to identify violation of the  monotonically  contractive characteristic feature
associated with  divisible Markovian mapping on the operator space.
We note that both the relative entropy and trace-distance  decline during  a complete positive  Markovian process, unlike the
  fidelity  which increases under the same conditions.
As pointed out earlier in Ref. \cite{usud}, negative values  of relative entropy difference (\ref{relD}) 
and the fidelity difference (\ref{fd}) only imply that the  time evolution is {\em not} a completely positive Markovian process, 
positive values of these same quantities does not imply the occurrence of a Markovian evolution.  Accordingly,
the negative values of relative entropy and fidelity differences serve as sufficient  but not necessary signatures of non-Markovianity (completely positive as well as non-completely positive). 
In order to highlight the latter point, we have repeated the calculations of the 
fidelity difference $G(t,\tau)$  (Eq.(\ref{fd})) as a function of $t, \tau$  done earlier, by using the
 trace-distance difference $D(t,\tau)$  (Eq.(\ref{traceD})
for the  cavity-cavity  two-qubit partition ( Eq.(\ref{tqubitm}) with
 the substitution $(1-p)\rightarrow q$), and illustrated   in Fig.~\ref{fideE}.

Comparing the results of Fig.~\ref{fideD} and Fig.~\ref{fideE}, we note that
non-Markovianity is optimized in the  vicinity of the exceptional point in both cases,
however there are subtle differences.The trace-distance difference measure yields
enhanced features in that it is able to detect the presence of non-Markovianity at small $\tau$
which remains undetected by the fidelity difference measure. Similar trends
have been noted \cite{usud} with the  relative entropy difference, which
shows the dependence of  non-Markovianity  dynamics on the metric entity.
It appears that  regions of contractive  quantum evolution vary according to the measure used to detect
violations of Markovian dynamics. In general one can conclude (at least on the basis of the
results obtained in this study) that these differences are marginal and
there is overall agreement in the critical regions of non-Markovianity.
In the next Section, we  make several important observations based on the
non-Markovian patterns of  Figs.~\ref{fideC}, \ref{fideD} and  \ref{fideE}
and results of non-locality computed using the  CHSH-Bell inequality.

\section{Violation of the CHSH-Bell inequality}\label{bellC}
\begin{figure}[htp]
  \begin{center}
    \subfigure{\label{fig6a}\includegraphics[width=4.0cm]{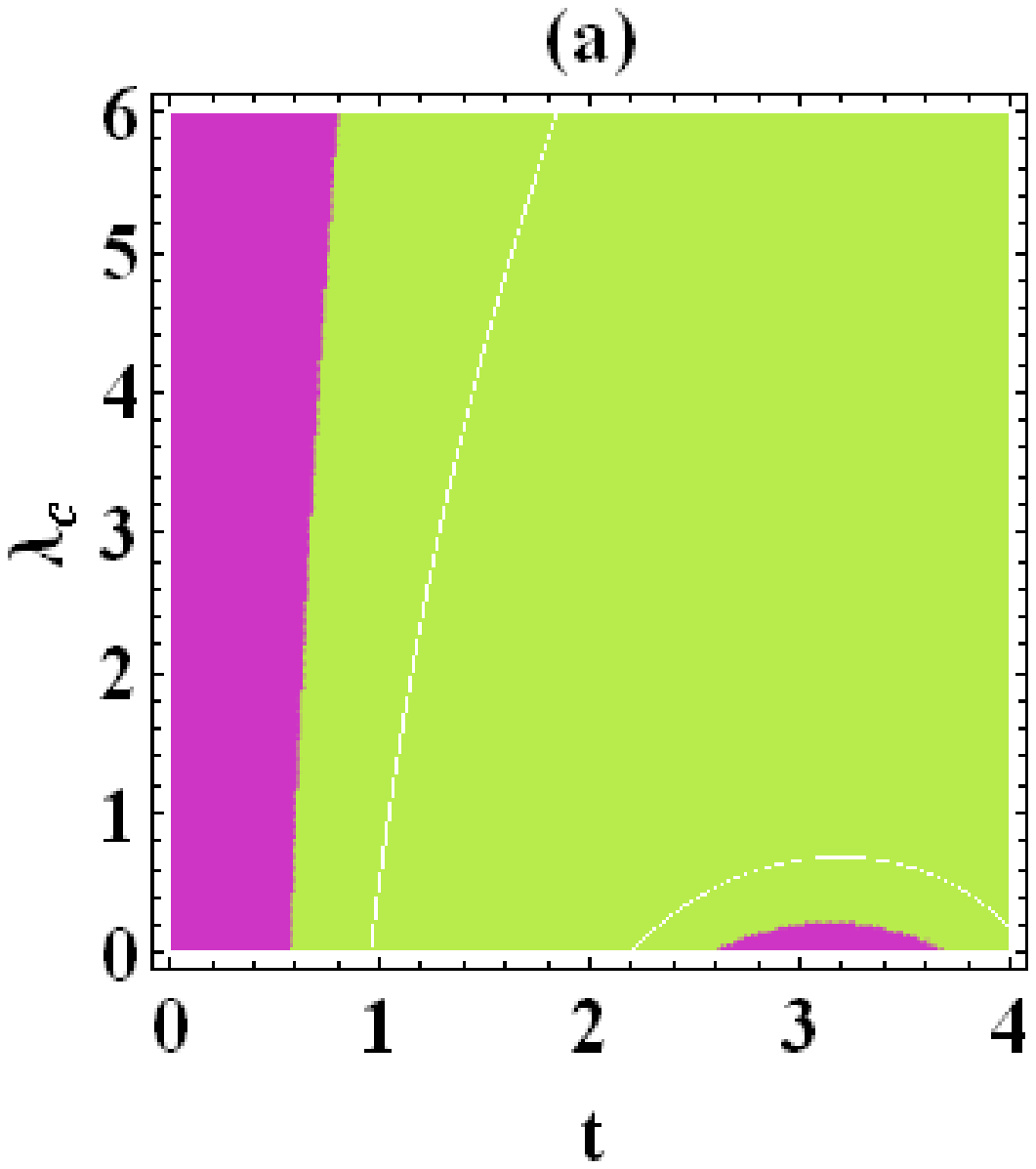}}\vspace{-1.1mm} \hspace{1.1mm}
     \subfigure{\label{fig6b}\includegraphics[width=4.0cm]{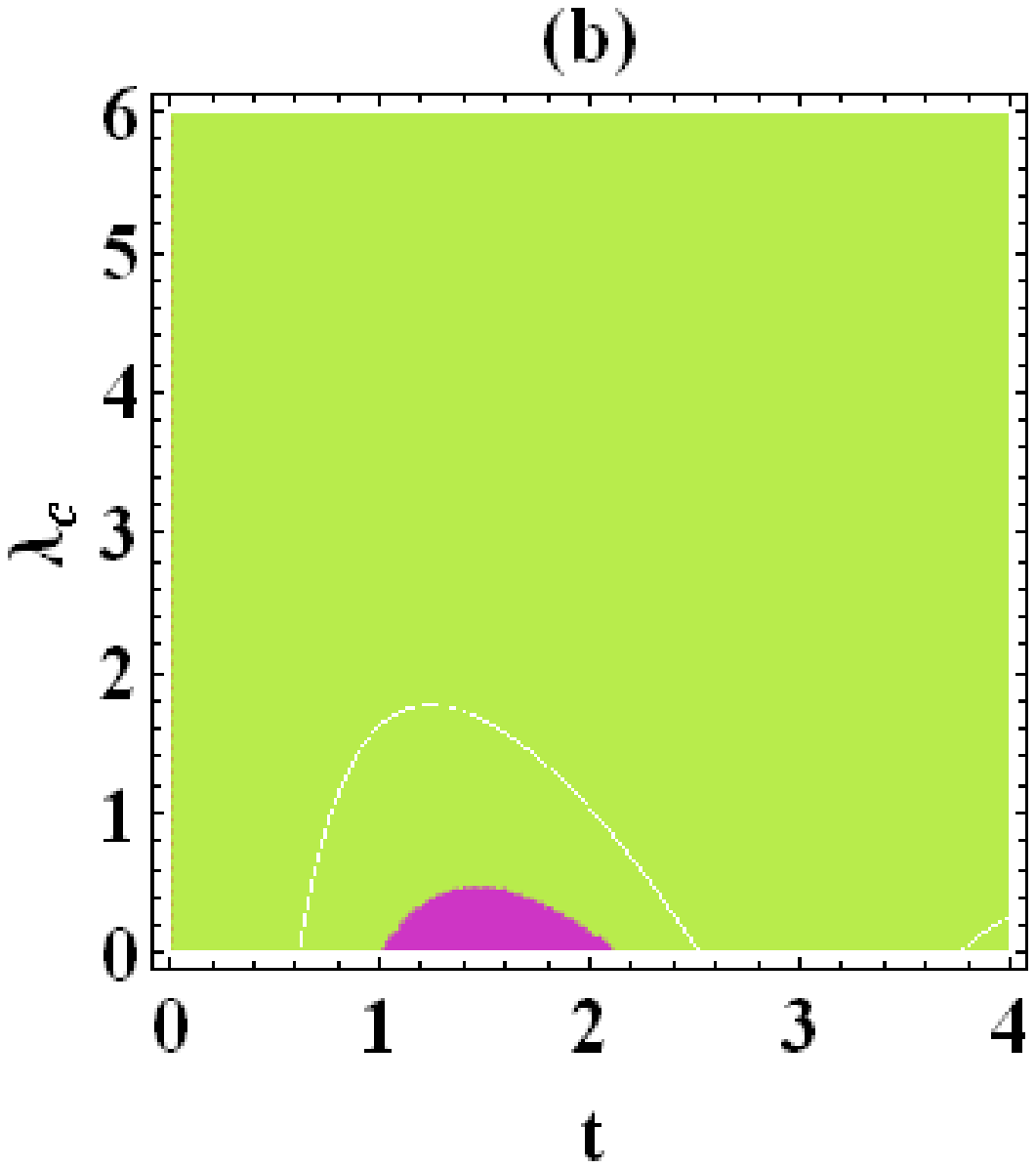}}\vspace{-1.1mm} \hspace{1.1mm}
 \subfigure{\label{fig6c}\includegraphics[width=4.0cm]{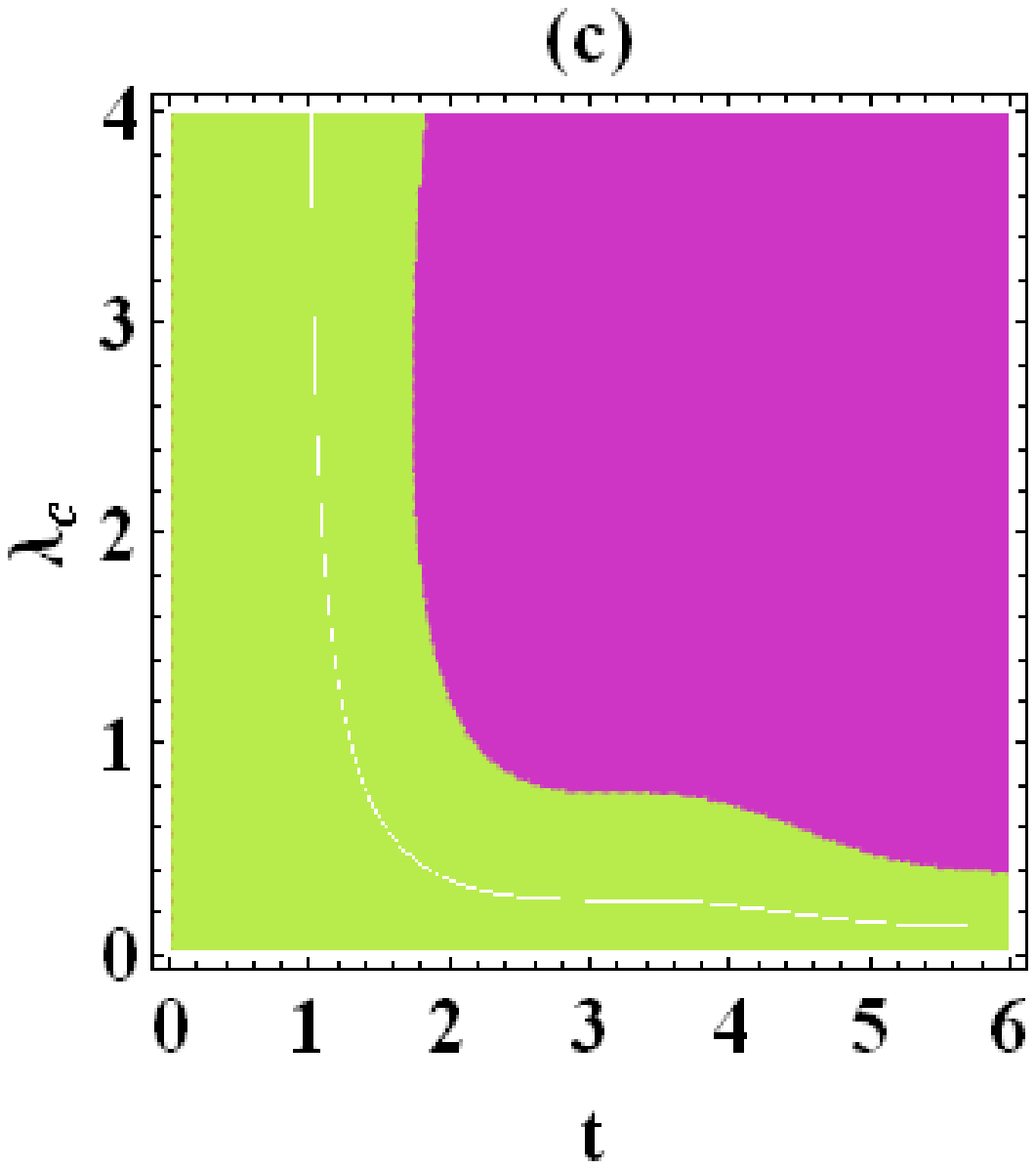}}\vspace{-1.1mm} \hspace{1.1mm}
 \subfigure{\label{fig6d}\includegraphics[width=4.0cm]{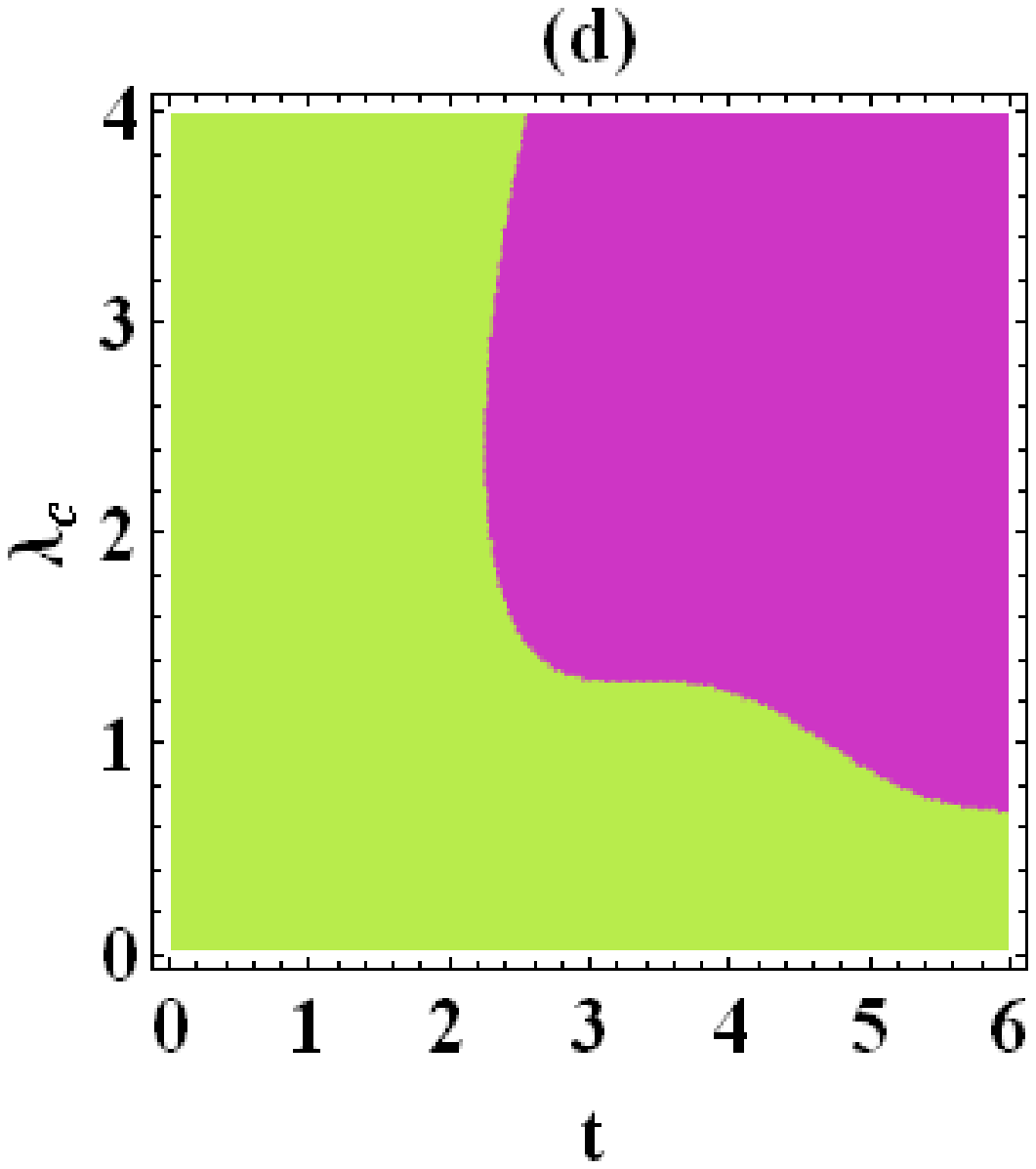}}\vspace{-1.1mm} \hspace{1.1mm}
     \end{center}
  \caption{Left to right:  $\mathcal{B}$
(a) for the atom-atom, (b) cavity-cavity,  (c) reservoir-reservoir and (d)
 inter-system atom-cavity  two-qubit partitions
  as a function of time $t$ and $\lambda_c$.
$a$=$b$=$\frac{1}{\sqrt 2}$. (a-c) and $a$=$\frac{1}{\sqrt 5}$ for (d).
 The remaining  intra-system atom-cavity, atom-reservoir and atom-cavity
partitions did  display any violation of the CHSH-Bell  inequality (using 
 Eq.(\ref{b1}).  Regions (shaded purple) for which  $\mathcal{B}$
exceeds 2 are  Bell nonlocal. The white lines divides the
 fine interplay between the two competitive terms   which appear in the Bell function
of Eq.(\ref{b1})}.
 \label{bello}
\end{figure}

The violation of the  Bell inequality $|\mathcal{B}| \leq 2$
is quantified by the CHSH- Bell  inequality function $\mathcal{B}$ of 
a  two-qubit density matrix  where
$\mathcal{B} = M(\vec{a},\vec{b})-M(\vec{a},\vec{b}')+M(\vec{a}',\vec{b})+
M(\vec{a}',\vec{b}')$.   $M(\vec{a},\vec{b})$ is the correlated 
results ($\pm 1$) arising from the measurement of two qubits  in 
directions $\vec{a}$ and $\vec{b}$. The CHSH-Bell inequality is violated when $\mathcal{B}$
exceeds 2, and the correlations is considered inaccessible by any classical means
of information transfer, while for  values less than 2, 
the local hidden-variable theory can satisfy the inequality.
 Here, we  investigate the role that 
the parameter $\lambda_c$ play in the  violation of a Bell inequality,
with view to seeking a link with the non-Markovianity measure evaluated for various partitions
in the earlier  section.
$\mathcal{B}$ is evaluated as a function 
of the  correlations averages and for a given  density matrix 
\be
\label{dmat}
 {\rho}=\left(
\begin{array}{cccc}
u_{11}& 0 & 0 & 0  \\
  0 & u_{22}&u_{23}& 0 \\
  0 &u*_{32}&u_{33}& 0 \\
   0& 0 & 0 & 0 \\
\end{array}
\right)\\
\ee
using simple relations\cite{bello}
\bea
\label{b1}
\mathcal{B}&=&  {\rm Max} \; \{\mathcal{B}_1,\mathcal{B}_2\} \\
\nonumber
\mathcal{B}_1&=& 2 \sqrt{4 |u_{23}|^2+(u_{11}-u_{22}-u_{33})^2 }\\
\nonumber
\mathcal{B}_2&=& 2 \sqrt{2 |u_{23}|^2}
\eea
Figure~\ref{bello} a,b,c,d  shows  the  regions (shaded purple) in which $\mathcal{B}$
exceeds 2,  for the atom-atom, cavity-cavity,  reservoir-reservoir  and inter-system atom-cavity  two-qubit
partitions. Interestingly, the intra-system atom-cavity, atom-reservoir and atom-cavity
partitions did not display any violation of the CHSH-Bell  inequality for the range of time $t$ 
and $\lambda_c$ considered in Figure~\ref{bello}. Figure~\ref{bello} c shows that
the reservoir-reservoir subsystem exhibits a larger region of violation 
of the CHSH-Bell  inequality when compared to all other subsystems. 

We can make several important observations by comparing
 the non-Markovian features of  Figs.~\ref{fideC}, \ref{fideD},  \ref{fideE}
and the  CHSH-Bell  inequality violation trends in Figure~\ref{bello} a,b,c,d.
It is evident  that non-locality
present in one subsystem (e.g  reservoir-reservoir subsystem)
 appears in conjunction with   non-Markovian dynamics  in an adjacent subsystem 
(e.g atom-reservoir or the cavity-cavity partition)  with respect to changes in time $t$ and decay rate 
$\lambda_c$.  The cavity-cavity partition,
for instance has notable non-Markovian features (Figs.~\ref{fideD},~\ref{fideE}),
yet displays non-locality only in  a narrow range of $t$ and $\lambda_c$.
 There is  also mismatch between non-locality and
non-Markovianity  in the inter-system atom-cavity  two-qubit partition,
with Bell nonlocal regions dominant at  large $\lambda_c > $ 1 at which the system dynamics is noted to be Markovian.
While  these trends seem applicable  to two-qubit partitions, it remains to be seen whether
higher dimensional qubit partitions follow similar trends.
Our final observations relates to the match between Bell non-locality and 
non-classical correlations for select partitions which we illustrate next.

\subsection{Classical and non-classical correlations}\label{disC}

In this section, we evaluate 
the correlation  measure known as the quantum discord, $\mathcal{D}$
for the two-qubit state in Eq.(\ref{tqubitm}) and its cavity-cavity and reservoir-reservoir
counterpart matrices. It is well known that the quantum discord 
 is more robust as it captures  nonlocal correlations not present in the  
entanglement measure \cite{zu,ve1,ve2}, consequently it is non vanishing
 in  states which has zero entanglement. The  evaluation of $\mathcal{D}$
involves lengthy optimization procedures and analytical expressions are known
 exist only in a few limiting cases \cite{qali,qali2,wangli}.

The quantum mutual information of a composite state $\rho$ of
 two subsystems $A$ and $B$  is  given by \cite{zu,ve1,ve2}
$\mathcal{I}(\rho) = S(\rho_A) + S(\rho_B) - S(\rho)$  for a density operator in $\mathcal{H}_A
\otimes \mathcal{H}_B$. $\rho_{A}$ ($\rho_B$) is the  
reduced density matrix associated with $A$ ($B$) and $S(\rho_i)$ (i=A,B) denotes 
the well known  von Neumann entropy of the density operator $\rho_i$, where
$S(\rho)= - {\rm tr}(\rho \log\rho)$. The mutual information  can be written as
\be
\label{d1}
\mathcal{I}(\rho) = S(\rho_B)  - S(\rho|\rho_A)
\ee
where $S(\rho|\rho_A)= S(\rho) - S(\rho_A)$ 
is  the quantum conditional entropy. 

After a series of measurements on $A$, the  post  measurement conditional state in $B$ is given by
$\rho_{B|k} = \frac{1}{p_k} (\Pi_k \otimes \mathbb{I}_B) \rho (\Pi_k \otimes
    \mathbb{I}_B)$ where the probability $p_k = {\rm tr}[\rho (\Pi_k\otimes \mathbb{I}_B)]$ and
  $\{\Pi_k\}$ denote the one-dimensional projector indexed by the 
outcome $k$.  A conditional entropy of the of the conditional state of  subsystem $B$, $\rho_{B|k}$,
is derived  based on the cumulative effect of the mutually exclusive measurements 
on $A$ as $S(\rho|\{\Pi_k\}) = \sum_k p_k S(\rho_{B|k})$.
The measurement induced mutual information is therefore given by
\be
\label{d2}
 \mathcal{I}(\rho|\{\Pi_k\}) = S(\rho_B) - S(\rho|\{\Pi_k\})
\ee
Eqs.(\ref{d1})  and (\ref{d2}) appear with slight differences due to incorporation of
measurements in the definition for the latter.

The  classical correlation measure based on optimal 
measurements is evaluated using
$ \mathcal{C}_{A}(\rho) = \sup_{\{\Pi_k\}} \mathcal{I}(\rho|\{\Pi_k\})$, and
is  the maximum
information about one subsystem that can be obtained via  measurements
performed on the adjacent subsystem. 
The quantum discord  which measures the non-classical correlations
is given by the difference in $\mathcal{I}(\rho)$ and
$\mathcal{C}_A(\rho)$
\be
\label{disc}
\mathcal{D}_{A}(\rho) =  
\mathcal{I}(\rho) - \mathcal{C}_A(\rho)
\ee

Following the analytical approach in earlier works \cite{qali,qali2,wangli} ,  expressions for the
$C$ and  $Q$ associated with the atom-atom bipartite system in  Eq.(\ref{tqubitm}) is obtained as
 \begin{eqnarray}
\label{ccorr}
\mathcal{C}({\rho}_{a_1,a_2}) &=&  {H}(|a|^2 (1-p)) \nonumber\\
&-& {H}\left (\frac12 \left [1+\sqrt{1-4|a|^2 (1-p) p} \right ] \right),
\\
\mathcal{D}(\rho_{a_1,a_2})&=& {H}(|b|^2 (1-p)) - H(1-p)\nonumber\\
&-& {H}\left (\frac12 \left [1+\sqrt{1-4|a|^2 (1-p) p} \right ] \right)
\label{qcorr}
\end{eqnarray}
 where the function  $H(x)=-x \log_2x-(1-x) \log_2(1-x)$.
Analogous expressions for $\mathcal{C}({\rho}_{c_1,c_2}), \mathcal{D}({\rho}_{c_1,c_2})$
and  $\mathcal{C}({\rho}_{r_1,r_2}), \mathcal{D}({\rho}_{r_1,r_2})$
 associated respectively with the 
cavity-cavity  and reservoir-reservoir density matrices at time $t$,
are obtained by respective substitutions $(1-p)\rightarrow q$ and $(1-p) \rightarrow \gamma_d$ 
in  Eq.~(\ref{ccorr}) and (\ref{qcorr}). At $t=$0 and $a=\frac{1}{\sqrt{2}}$, 
$\mathcal{C}({\rho}_{a_1,a_2})$=$\mathcal{C}({\rho}_{a_1,a_2})$=1 while the correlations
of the cavity-cavity  and reservoir-reservoir partitions are zero.
Figures for the classical and quantum correlations as function of time $t$ 
for the analytical solutions ($p,q$) in  Eqs.(\ref{co1}),  (\ref{co2}) are 
shown  below.

\begin{figure}[htp]
  \begin{center}
    \subfigure{\label{figCa}\includegraphics[width=4.4cm]{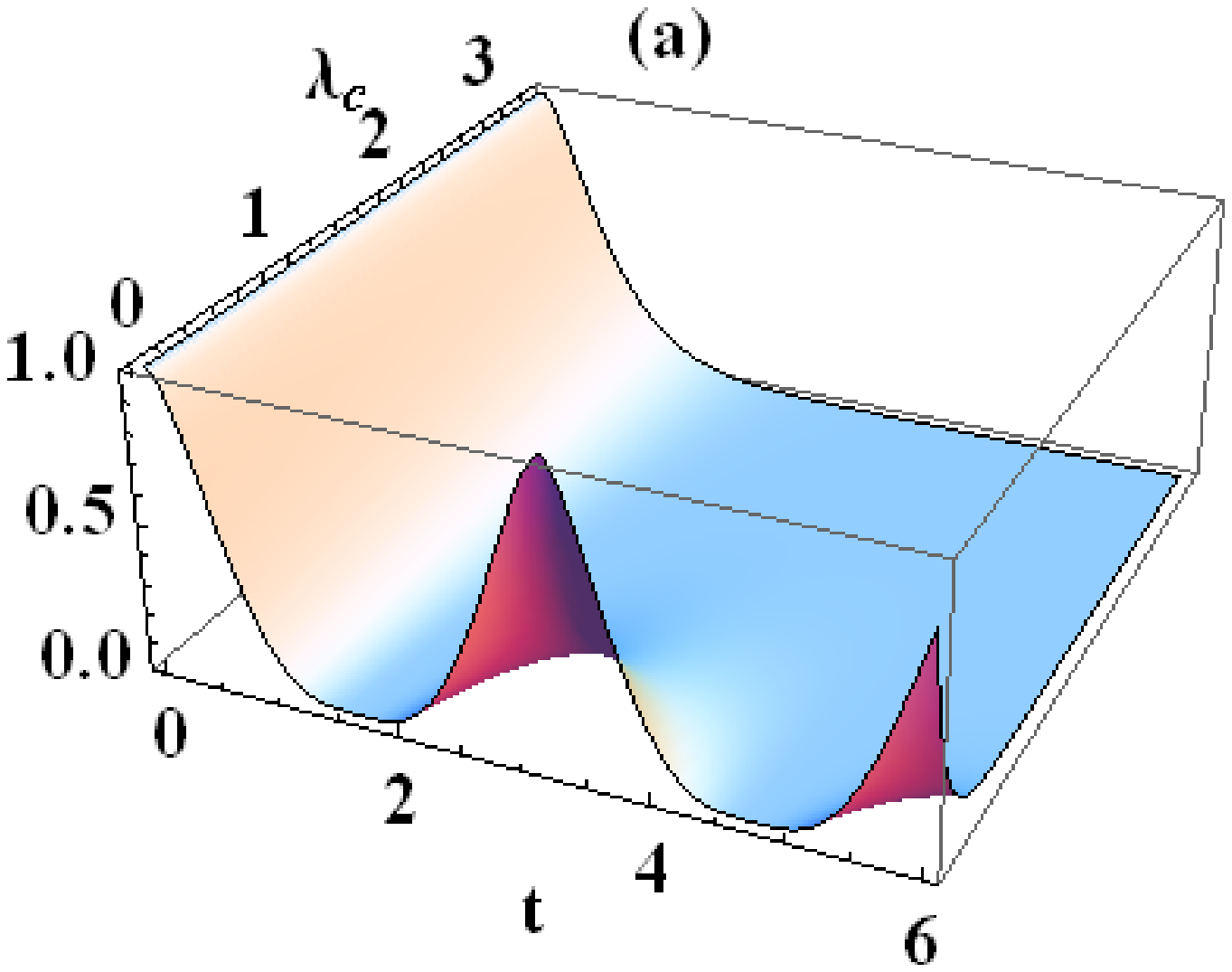}}\vspace{-1.1mm} \hspace{1.1mm}
     \subfigure{\label{figCb}\includegraphics[width=4.4cm]{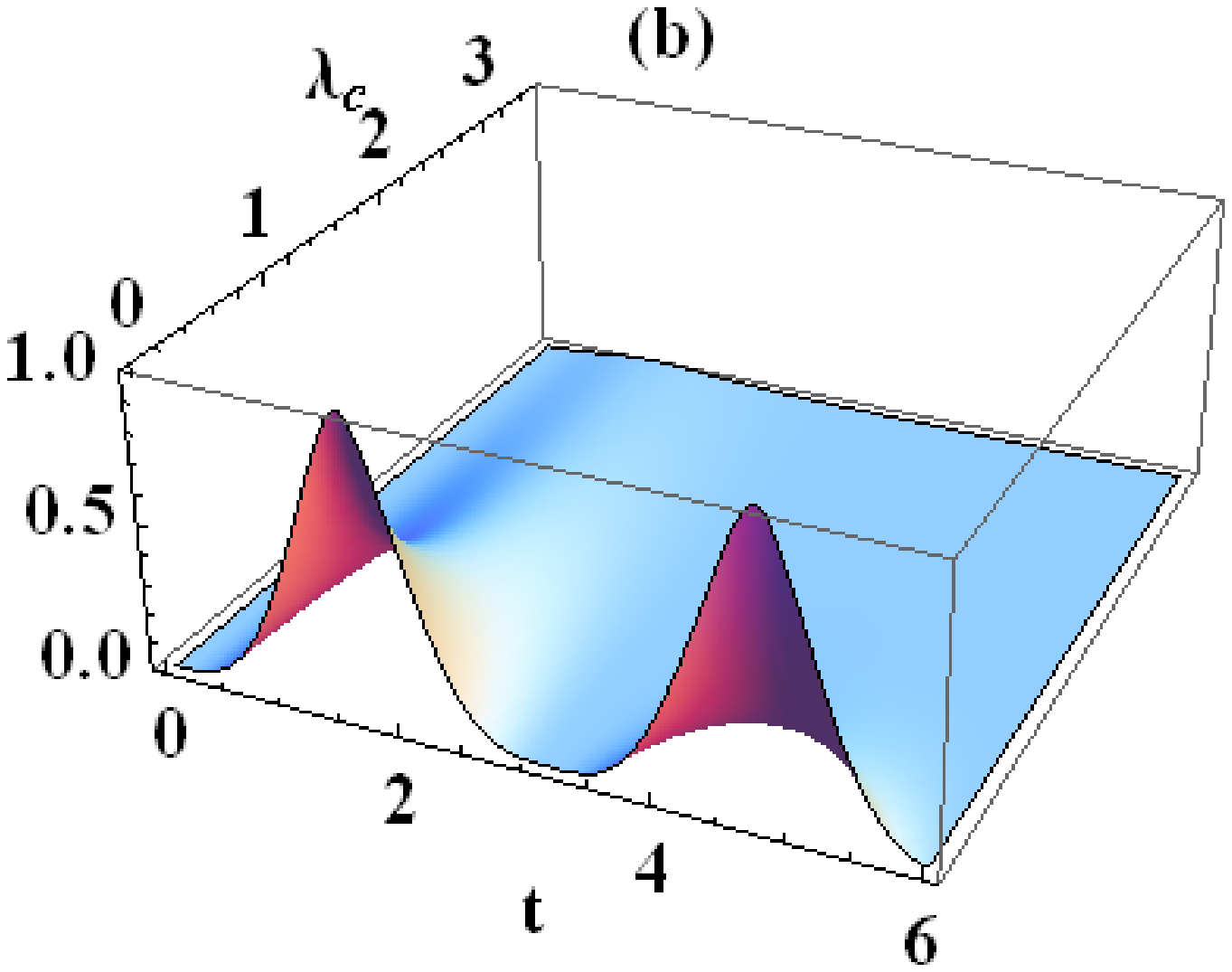}}\vspace{-1.1mm} \hspace{1.1mm}
 \subfigure{\label{figCc}\includegraphics[width=4.4cm]{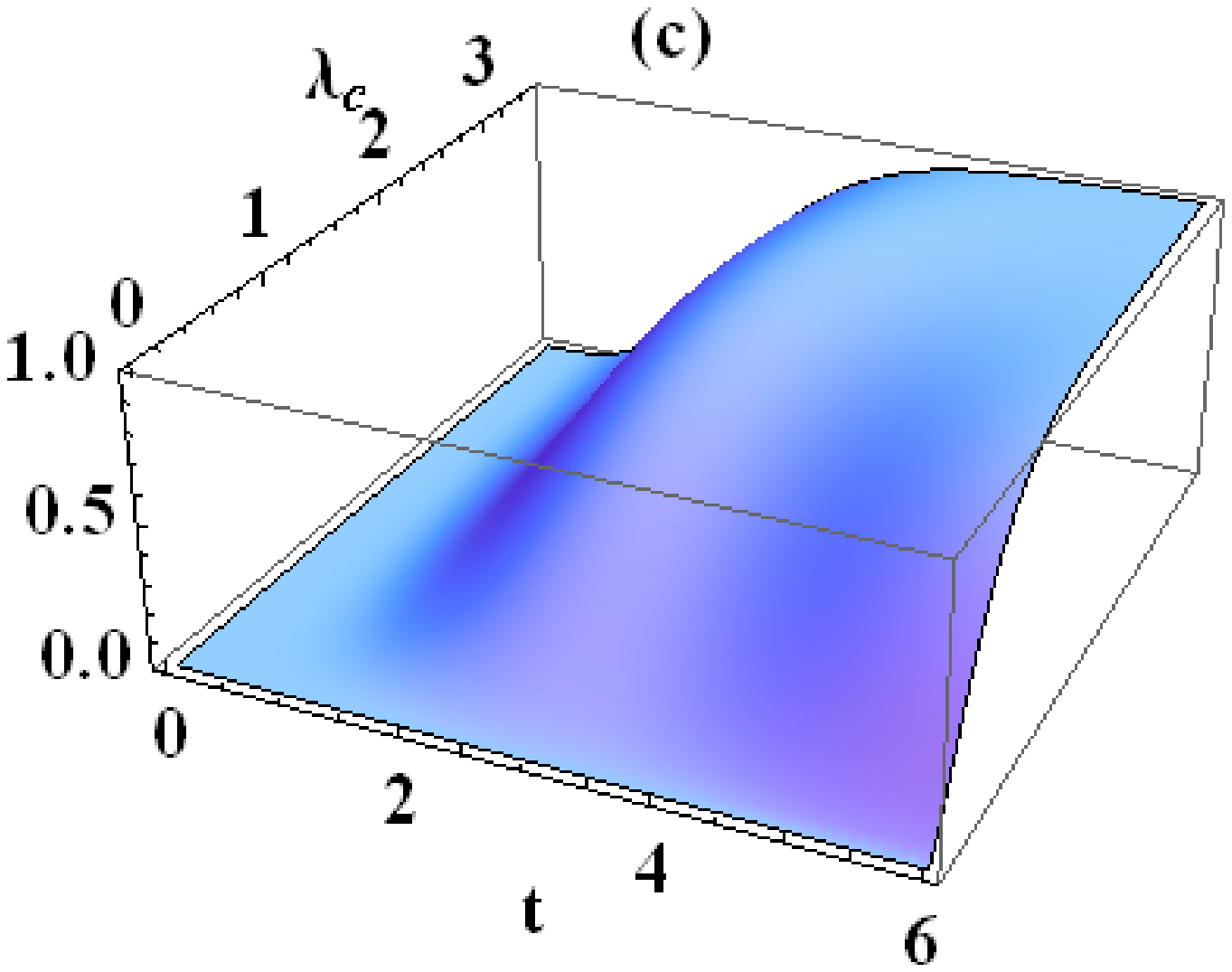}}\vspace{-1.1mm} \hspace{1.1mm}
     \end{center}
  \caption{Classical correlation $\mathcal{C}({\rho})$  for the (a)  atom-atom, (b) cavity-cavity,  (c) reservoir-reservoir 
  two-qubit partitions as a function of time $t$ and $\lambda_c$.  $a$=$b$=$\frac{1}{\sqrt 2}$.
}
 \label{class}
\end{figure}

\begin{figure}[htp]
  \begin{center}
    \subfigure{\label{figCa}\includegraphics[width=4.4cm]{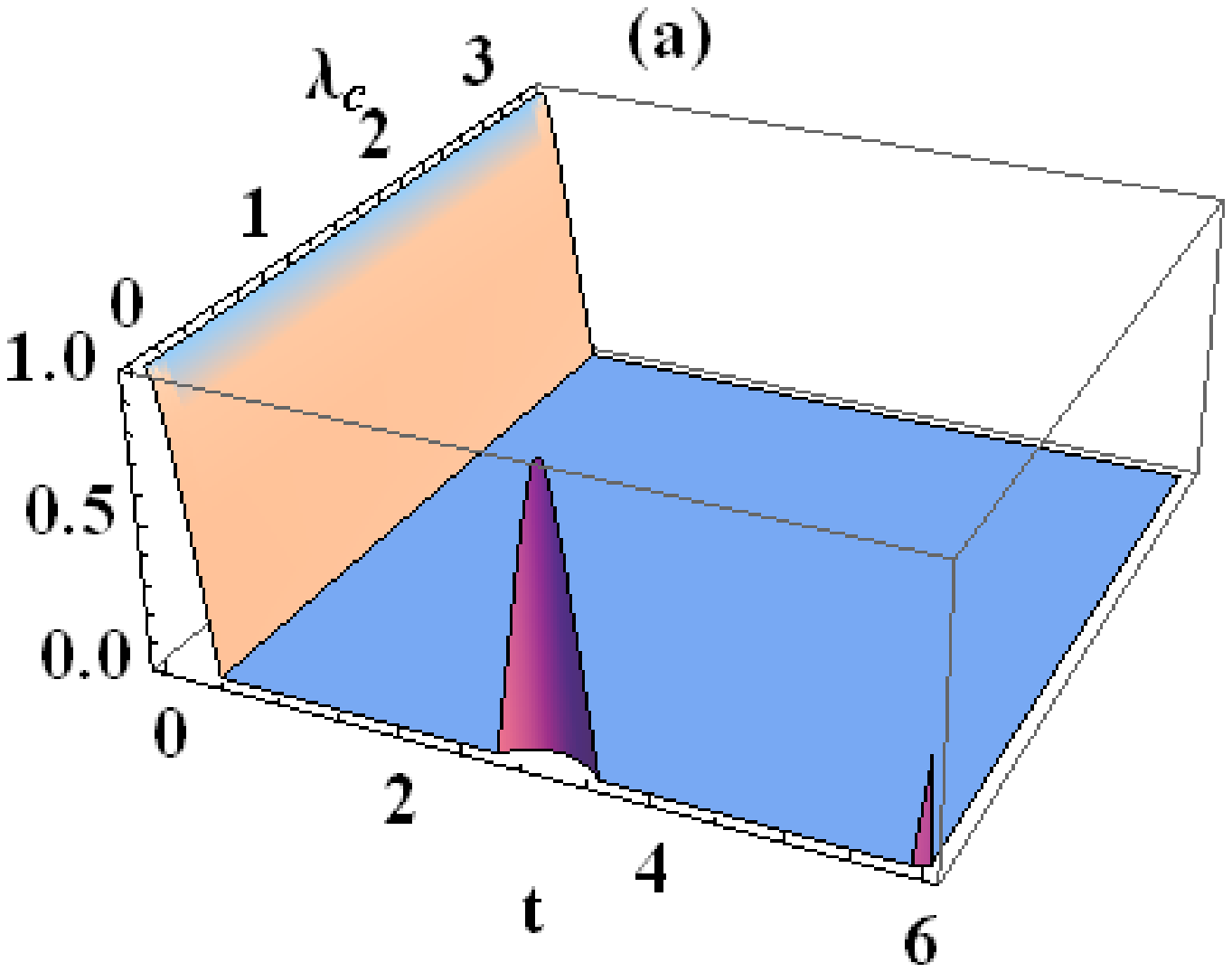}}\vspace{-1.1mm} \hspace{1.1mm}
     \subfigure{\label{figCb}\includegraphics[width=4.4cm]{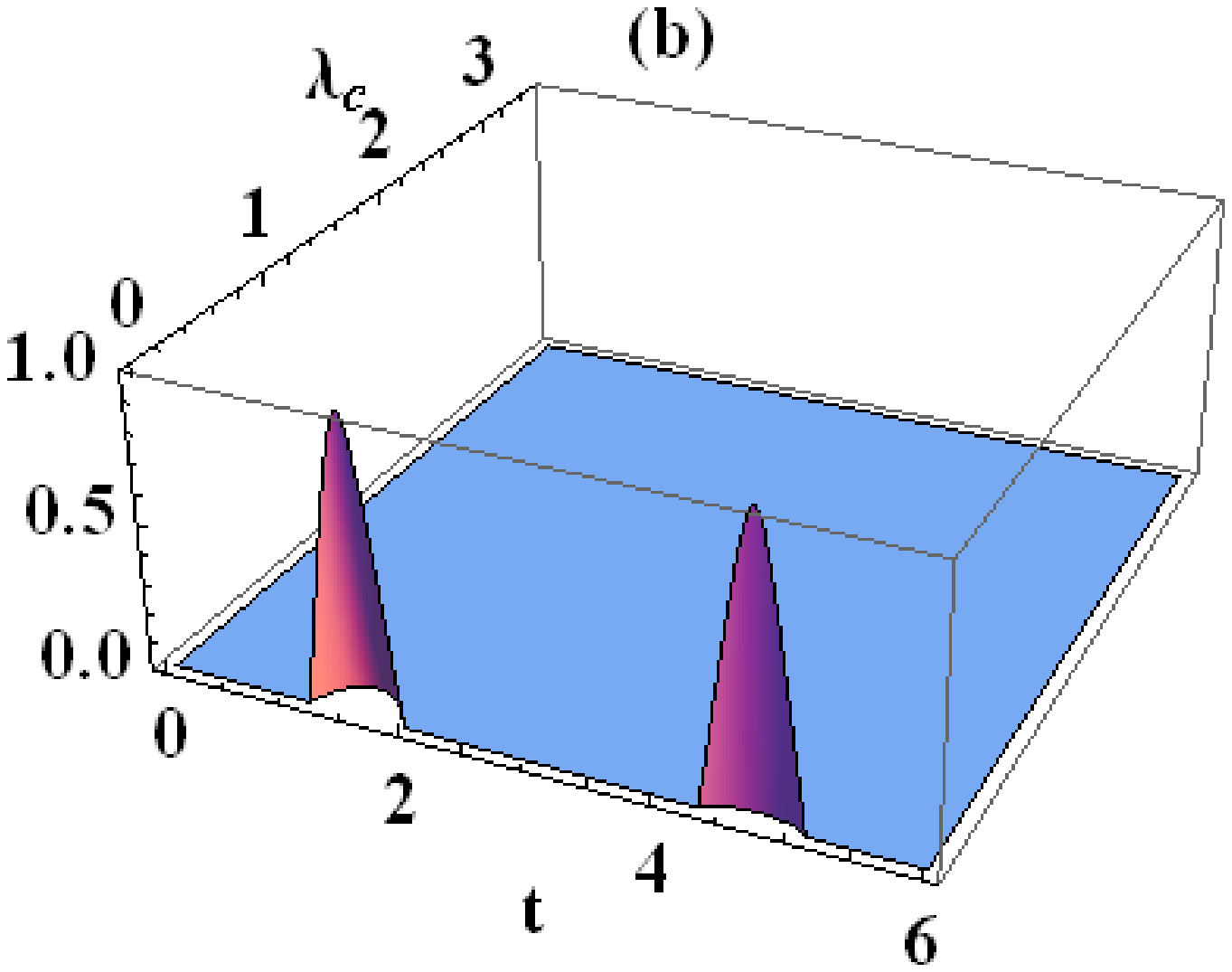}}\vspace{-1.1mm} \hspace{1.1mm}
 \subfigure{\label{figCc}\includegraphics[width=4.4cm]{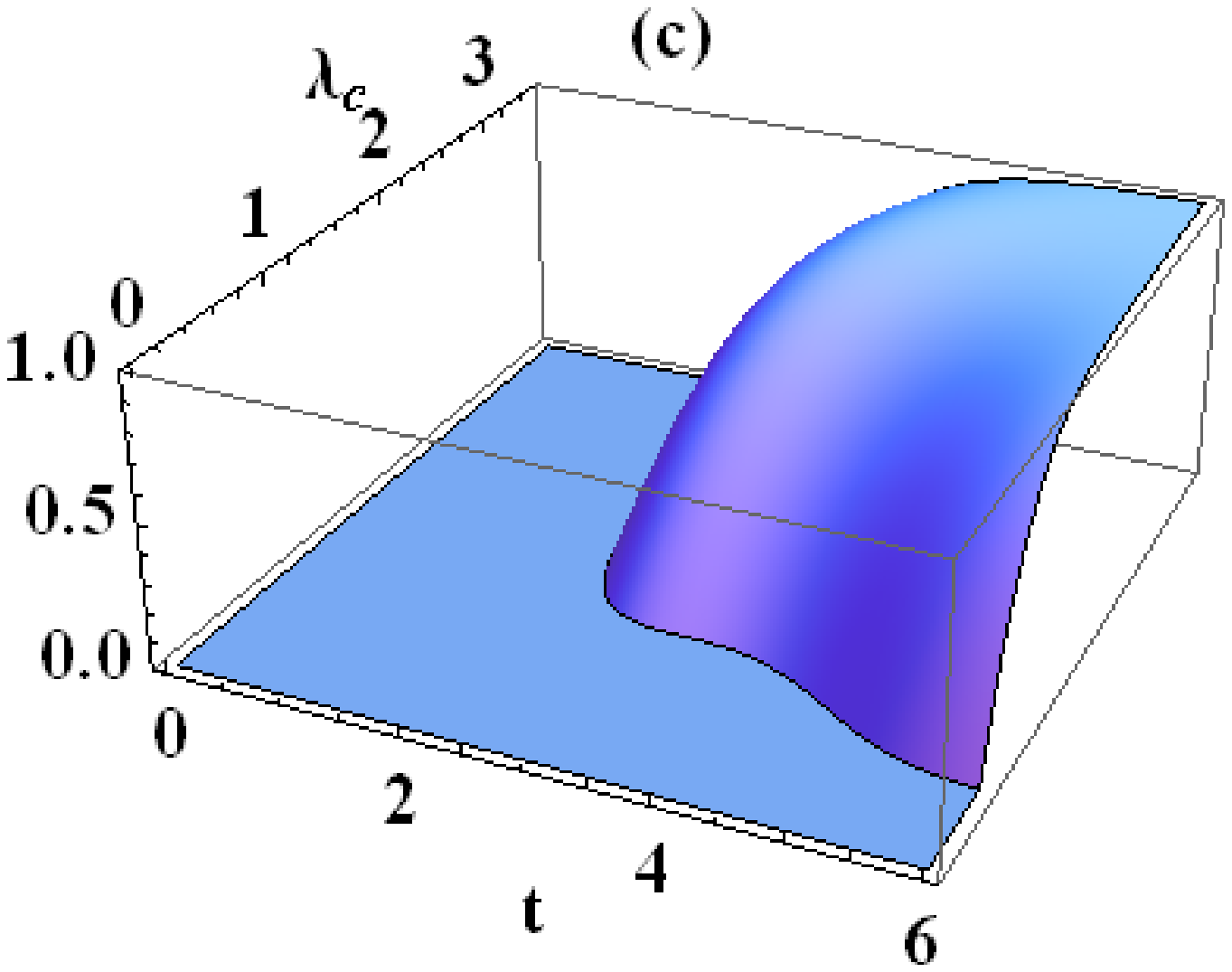}}\vspace{-1.1mm} \hspace{1.1mm}
     \end{center}
  \caption{Quantum correlation $\mathcal{D}({\rho})$  for the (a)  atom-atom, (b) cavity-cavity,  (c) reservoir-reservoir 
  two-qubit partitions as a function of time $t$ and $\lambda_c$.  $a$=$b$=$\frac{1}{\sqrt 2}$.
There is transfer of the quantum discord
from the atom-atom  subsystem to the reservoir-reservoir partition.}
 \label{dis}
\end{figure}
Figures~\ref{class} and ~\ref{dis} show the expected shift in classical $\mathcal{C}({\rho})$ and quantum correlations $\mathcal{D}({\rho})$ away from the entangled atom-atom partition to the reservoir-reservoir partition. There exist larger regions
 (as function of time $t$ and $\lambda_c$)  with  zero quantum correlations than classical correlations.
The appearance of  $\mathcal{D}({\rho})$ is most significant in the reservoir-reservoir 
  two-qubit partition as the system nears the exceptional point $\lambda_c \rightarrow$ 4.
Comparison of  Figure~\ref{bello} a,b,c and Figures~\ref{class} and
~\ref{dis}, show   that for the considered subsystem partitions 
(atom-atom,  cavity-cavity, reservoir-reservoir), Bell nonlocal regions
can be matched with appearance of non-trivial classical and non-classical correlations.
This is most noticeable in the reservoir-reservoir subsystem, and once again we reiterate
that these results appear valid for the two-qubit partitioned subsystems.

\subsection{Phase space approach of  quantum dissipation}\label{phase}

The current study has highlighted some salient features  linking entities 
based on the abstract Hilbert space , and within the confines of the two-qubit partitions
examined in this work. However, the validity of the results of dynamics in various subsystems 
is expected to be dependent on the form of the quantum master equation  utilized in Eq.~(\ref{pme2}).
Possible artifacts which arise from such Markovian forms
have  been examined by Kohen $et$ $al$.  \cite{kohen},
who analysed  time evolving density operators transformed 
using several approaches including  Redfield theories, the master equations of Agarwal \cite{agarwal} and 
 the semigroup theory of Lindblad \cite{lind}, into  Wigner phase space distribution functions.
The Agarwal-Redfield (AR) equations of motion were seen  to violate the semigroup form of Lindblad,
showing  that for certain initial states, the  Redfield theory can violate simple positivity
 making it difficult to assign distinctly physical status for such states.

The Agarwal bath \cite{agarwal} which is 
 translational invariant and able to reach the thermal equilibrium limit,
 yields density matrix positivity
for many initial conditions. However this bath system
presents pathological   effects as was noted in the density matrix negativity
 for a range of initial conditions at low temperatures \cite{talkner}
in a model system of one primary oscillator. These
observations suggest possible association between the 
 density matrix positivity and conditions which lead to 
the  thermal equilibrium in system-bath correlations. In this regard, there is
need to examine these links with incorporation of detailed attributes of the  dissipative sink inherent
in  the  the qubit-cavity-reservoir system of  Eq.~(\ref{ham}),  in future  investigations.

Finally, it is not entirely clear as to the role played by the bipartite  partitioning of the
six-qubit multipartite system  (Eq.~(\ref{mut6})) in the observed matching or non-matching
of regions with noticeable  non-locality, non-Markovianity
and quantum correlations in the subsystems examined in this study.
Future investigations  should be aimed at higher dimensional density matrices (e.g. tripartite systems)
and among all possible combinations of partitions, to see if the trends observed in two-qubit systems
are indeed universal. This can be a formidable task, which may require 
 a rigorous approach \cite{anand,eis} involving the mathematical formulations of the underlying
abstract Hilbert space,  to  help   form a basis for understanding
the links between the  quantum entities  such as non-locality and 
non-Markovianity.

\section{Application to Light-harvesting systems}\label{lhs}

The results of this study has immediate  relevance to
photosynthetic energy transfer in 
light-harvesting  systems such as the Fenna-Matthews-Olson complex
 with moderate exciton-phonon coupling, and operation under physiological conditions.
There is obvious similarity between the qubit-cavity-reservoir system in  Eq.~(\ref{ham})
and a photosynthetic  model  that constitutes the donor and acceptor protein pigment complexes with
finite energy difference and a third   party  with a continuous frequency spectrum acting as the
phonon-dissipative sink, and  into which energy from the acceptor dissipates with time.
The  inclusion of comprehensive details incorporating the
 solvent relaxation rate as well as  kinetic behaviors
of the electron transfer mechanisms and dynamic bath effects in 
 Eqs.~(\ref{pme4}) and ~(\ref{pme5}) will extend the applicability of the model in  Eq.~(\ref{pme2})
to the more realistic environment of photosynthetic systems. 

Earlier studies \cite{carupra,caru1}  have shown that attributes such as the non-
Markovian interactions and quantum correlations contribute 
via a delicate interplay of quantum mechanical attributes and the environmental noise 
 in bringing about an optimal  performance.
 of the light harvesting complex. 
Non-Markovian effects  were seen to reduce the transport efficiency while
increasing the lifetime of entanglement \cite{carupra} 
by disrupting the optimal balance of quantum and incoherent dynamics
required for efficient energy transfer. 
In another study involving light harvesting systems  \cite{reben2}, 
non-Markovian processes were noted to be 
 most pronounced  in the reorganization energy regime
where the transfer efficiency is the highest,  and linked with the 
preservation of coherence along critical pathways \cite{reben}.

Results based on  the exciton entanglement dynamics of the 
 Fenna-Matthews-Olson (FMO) pigment-protein  complex \cite{thilchem2},
indicate increased oscillations of entanglement  in the non-Markovian
regime, with implications for a link between non-Markovianity
and large coherence times.  Interestingly,  an earlier work on optimizing energy transfer efficiency
  by Silbey and coworkers \cite{silbey}, 
showed  that the interplay of coherent dynamics and
environmental noise resulted in the optimal energy transfer efficiency  at an intermediate level for several variables
in light-harvesting systems. In particular, the reorganization energy and the bath relaxation rate played critical roles 
in displaying a non-monotonic-dependence on energy transfer.
Spatial correlation effects were noted \cite{silbey} to  optimize the energy transfer only  during
strong  dissipation,  highlighting  differences in the roles played by spatial
correlation and temporal correlations. In this regard, there is need to further 
examine the subtle links between non-Markovian dynamics
and the effective  system-bath coupling range, 
 taking into consideration the site energy  based electronic
coupling correlations \cite{recom}.
Future investigations along these lines can employ the model developed through
 Eq.~(\ref{pme2}), (\ref{pme4}) and ~(\ref{pme5}) as
currently,  the information on the exact mechanism by which 
non-Markovian processes act to  bring  about fast energy transport  still remains unclear
in light-harvesting systems. 

The inclusion of quantum effects involving non-locality and non-classical correlations
will  provide greater insight to the recently
 proposed   optical cavity quantum electrodynamics  setup  \cite{caruexpt} 
to investigate  energy transfer mechanism in 
 biomolecules. There is also scope  to examine similar quantum effects in 
 a viable Grover-like search process \cite{grover} via the exciton trapping mechanism at dissipative
sink sites  in photosynthetic systems.
The incorporation of  an approach
incorporating non-Markovianity and the violation of the CHSH-Bell inequality
may assist in  developing new perspectives in interpreting experimental
results obtained via  two-dimensional electronic spectroscopy \cite{caram},
 4-wave mixing measurements \cite{segale} and
the hybrid optical detector \cite{tomo} which implements quantum detector tomography
and characterizes  simultaneous wave and photon-number sensitivities.
To this end, future experimental work should incorporate the  joint
extraction of  non-Markovian and Bell non-local features in biochemical systems,
pending  further refinement in optical techniques.

\section{Discussion and Conclusion}\label{con}

In summary, we have studied the links between non-locality, non-Markovianity
and quantum correlations of two  initially  correlated  atomic qubits, 
each located in a single-mode leaky cavity and  interacting with its own bosonic reservoir.
The appearance of  non-Markovianity is associated with a negative value for the
the  fidelity  or trace-distance  differences, and act only as  qualitative
signatures  of deviation from a complete positve Markovian behavior.
Results of the differences in non-Markovianity based on the two measures:
fidelity and trace-distance shows its  dependence on the distance metric,
and regions of contractive  quantum evolution vary according to the defined measure used to detect
violations of Markovian dynamics.  

The non-Markovian features in different two-qubit partitions  (cavity-cavity and
atom-reservoir partitions of the same subsystem,
 cavity-reservoir partition  across  different subsystem) show varying levels of
erasure or enhancement of non-Markovianity  depending on the cavity
 decay rate, $\lambda_c$. In particular, the  exceptional point
at which two eigenvalues merge, appear to signal the enhancement
of non-Markovianity in selected regions (of time $t$ and $\lambda_c$)
for the cavity-cavity partition. The changes in the vicinity of
the exceptional point, also noted in quantum measurements \cite{thilaIOP}, is significant 
as it may assist in distinguishing
 between processes with a classical origin and those with an intrinsically
quantum mechanical coherence origin \cite{miller}.

Results of the CHSH-Bell  inequality computed
for various two-qubit partitions  show  that enhanced non-locality
present in a specific subsystem (e.g  reservoir-reservoir matrix) appears
in conjunction  with  non-Markovian evolution in an adjacent subsystem 
(e.g atom-reservoir matrix). The mismatch between non-locality and
non-Markovianity for a given partition is in contrast to the matching between non-locality
and quantum correlations for regions spanned by time $t$ and  the cavity
 decay rate, $\lambda_c$. These  results appear to provide weak signatures of subtle links between non-Markovianity,
non-locality and non-classical correlations, bearing in mind that attempts to
establish such links are very much reliant  on the rigid formulation of the 
hidden variable model via the CHSH-Bell  inequality relations in  Eq.~(\ref{b1}).
Moreover  the  criteria used in this study provides  only sufficient and not necessary tests of non-Markovianity,
and also very much dependent on the initial state of correlations of the various subsystems.
Currently, it is still not clear as to the origins of non-Markovianity as there is uncertainty as to whether
information back-flow occurs due to the  initial system-environment correlations or 
the intrinsic presence of specific dynamical features from which non-Markovianity emerge.
Moreover,  investigations involving tripartite states 
and higher dimensional states is needed to seek a definite conclusion on the perceived
correlation/anti-correlation between non-locality, non-Markovianity and 
quantum discord. Lastly, the results obtained in this work highlight the 
need to adopt a unified model that  incorporates non-Markovianity, 
non-locality and non-classical correlations, as an effective approach
 in examining high  efficiencies of  energy transfer observed  in 
 light-harvesting systems.

\section*{Acknowledgments}
The authors gratefully acknowledge  useful comments from
Prof. Rajagopal (Inspire Institute Inc., Virginia,  USA) and the anonymous referees.
This research was undertaken on the NCI National Facility in Canberra, Australia, which is supported by the Australian Commonwealth Government.

\end{document}